\documentclass[journal]{IEEEtran}
\usepackage{cite}
\usepackage{hyperref}

\usepackage[cmex10]{amsmath}
\usepackage{xcolor}

\usepackage{amsfonts}
\usepackage{amssymb}
\usepackage{array}
\usepackage{graphicx}
\usepackage{subfigure}
\newtheorem{lem}{Lemma}
\newtheorem{thm}{Theorem}
\newtheorem{cor}{Corollary}
\newtheorem{definition}{Definition}
\newtheorem{assum}{Assumption}
\newtheorem{rem}{Remark}

\def\endprf{{$\hfill\square$\\}}

\usepackage{graphicx}          

\allowdisplaybreaks

\def\endprf{{$\hfill\square$}}
\begin{document}

\title{Distributed Resource Allocation over Time-varying Balanced Digraphs with Discrete-time Communication}


\author{Lanlan~Su, Mengmou~Li, Vijay~Gupta, and Graziano~Chesi%
\thanks{L. Su is with School of Engineering, University of Leicester, Leicester, LE1 7RH, UK (email: ls499@leicester.ac.uk); M. Li and G. Chesi are with the Department
of Electrical and Electronic Engineering, University of Hong Kong, Hong Kong (email:mmli.research@gmail.com, chesi@eee.hku.hk); V. Gupta is with Department of Electrical Engineering,
University of Notre Dame, Notre Dame, IN 46556 USA (email:vgupta2@nd.edu).}}%


\maketitle

\begin{abstract}  
This work is concerned with the problem of distributed resource allocation in continuous-time setting but with discrete-time communication over infinitely jointly connected and balanced digraphs. We provide a passivity-based perspective for the continuous-time algorithm, based on which an intermittent communication scheme is developed. Particularly, a periodic communication scheme is first derived through analyzing the passivity degradation over output sampling of the distributed dynamics at each node. Then, an asynchronous distributed event-triggered scheme is further developed. The sampled-based event-triggered communication scheme is exempt from Zeno behavior as the minimum inter-event time is lower bounded by the sampling period. The parameters in the proposed algorithm rely only on local information of each individual nodes, which can be designed in a truly distributed fashion.
\end{abstract}

\section{Introduction}
An important distributed optimization problem is one in which each node has access to a convex local cost function, and all the nodes collectively seek to minimize the sum of all the local cost functions \cite{nedic2009distributed,zhu2011distributed,gharesifard2013distributed,shi2015extra}.
Most optimization algorithms reported in the literature are implemented in discrete time. 
However, as pointed out by \cite{wang2011control,yang2019survey},  discrete-time algorithms might be insufficient  for applications where the optimization algorithm is not run digitally, but rather via the dynamics of a physical system,  such as collectively optimizing social, biological and natural systems, robotic systems \cite{zhao2017distributed}.  \textcolor {black}{The resource allocation, as an important class of distributed optimization problem, has been studied in continuous-time setting \cite{cherukuri2015distributed,yi2016initialization, deng2017distributed,kia2017distributed,ding2018distributed, zhu2019distributed} and discrete-time setting \cite{li2015distributed,shi2018distributed,doan2020distributed}. The existing works concerned with the distributed resource allocation problem assume topology graphs to be fixed over time and/or do not take the communication cost into account. In this work, we aim at providing a passivity-based perspective for a continuous-time algorithm of  distributed resource allocation over time-varying digraphs, based on which an intermittent communication scheme is developed. }

\textcolor{black} {Passivity serves as a useful tool to analyze multi-agent systems (MASs).  This has been well illustrated  by  \cite{chopra2006passivity} which shows that a MAS of  possibly heterogeneous  agents can reach consensus over a time-varying balanced strongly connected graph as long as all individual agents are  input-output passive. In \cite{li2019consensus}, we generalize the results of \cite{chopra2006passivity} to MASs with all agents that can be characterized by a passivity index.  The current work is rooted in the same idea, but we would like to note that it is not trivial to apply the idea for consensus of MASs to distributed optimization problems. One of the main challenges  is to verify that the individual algorithmic dynamics of a particular distributed algorithm is dissipative with a  quantified input feed-forward passive (IFP) index. On the other hand, the passivity and dissipativity (including IFP as a special case) have been recently exploited in networked control systems coping with different communication imperfections. For instance, \cite{yu2013output} addresses the problem of output synchronization of passive systems with event-triggered communication wherein network delay and quantization are considered as well; in \cite{yan2019analysis},  passivity index has been  used to feedback control two-dimensional systems over digital communication network wherein output sampling and event-triggered scheme are designed; \cite{lee2014passivity} uses a passivity framework to model and mitigate attacks on networked control system; \cite{wang2014feedback} considers the packet drops of the communication channel. See \cite{zakeri2019recent} for more recent works on cyber-physical systems using passivity indices. This forges the main motivation for us to provide a passivity-based perspective for the algorithm as the IFP framework opens up the new possibilities of implementing the algorithm over a nonideal digital communication network and reduce the channel usage.}

\textcolor{black}{ Whereas there exist considerable works on algorithms design for distributed optimization, there are relatively fewer works that take communication constraints into account. In this work, we are interested in intermittent communication including periodic (also called sampled-based) and event-triggered manners. The underpinning distributed algorithms proposed in the existing works in this direction are either discrete-time \cite{li2015distributed, lu2017event, kajiyama2018distributed, shi2018distributed, liu2019distributed, liu2020resource} (see Tables 1 and 2 in \cite{yang2019survey} for a comprehensive list) or continuous-time \cite{kia2015distributed, chen2016event, yu2017distributed, du2018distributed,wang2018event, yi2018distributed,ding2018distributed, liu2019event,shi2020distributed} (see Table 3 in \cite{yang2019survey} for a comprehensive list). Among the works concerned with continuous-time algorithms with a discrete-time communication scheme, the results in \cite{ shi2020distributed, yu2017distributed, du2018distributed,yi2018distributed,kia2015distributed, ding2018distributed} are limited to undirected and fixed topological graphs while \cite{chen2016event,wang2018event,liu2016event} assume the graph to be strongly connected and fixed over time. \cite{liu2019event} studies the problem of event-triggered distributed optimization over a uniformly jointly connected graph but restricts it to be undirected. } To the best of our knowledge, distributed optimization problem over time-varying jointly connected digraphs has never been addressed in the continuous-time setting due to the difficulties from stability analysis under the time-varying nature and lack of connectedness of topologies. \textcolor{black}{The passivity-based method has been shown to be powerful in handling communication imperfections and in distributed control, thus being a promising approach to treat both the time-varying graph topology and communication.}

\textcolor{black}{In this work, we consider the problem of distributed resource allocation over a time-varying digraph under intermittent communication.} Specifically, each node has access to its own local cost function and local network resource, and the goal is to minimize the sum of the local cost functions subject to a global network resource constraint. \textcolor{black}{The communication topology is described by a weight-balanced and infinitely jointly strongly connected digraph.} Closest papers which have also exploited the notion of passivity to address the distributed optimization problem are  \cite{tang2016distributed,hatanaka2018passivity}. The results in these mentioned works are limited to  fixed and undirected  graphs. \textcolor{black}{Our work features a novel passivity-based perspective for continuous-time algorithms, which enables us to design an  intermittent communication scheme  over infinitely jointly strongly connected digraphs.  Starting from a continuous-time algorithm, a periodic communication scheme is first derived  through  analyzing the passivity degradation over output sampling of the distributed dynamics at each node. Then, an asynchronous distributed event-triggered  scheme  is further developed. The sampled-based event-triggered communication scheme is exempt from Zeno behavior as the minimum inter-event time is lower bounded by the sampling period. The parameters in the proposed  algorithm rely only on local information of each individual nodes, which can be designed in a truly distributed fashion.}

\vspace{2mm} 
 \textcolor{black}{The rest of this paper is organized as follows.   Section \ref{sec:2} introduces some preliminaries and states the problem formulation. Section \ref{sec:3} presents the main results. Specifically, Section \ref{sub:The-Lagrange-dual} reformulates the problem into its dual distributed convex optimization problem. Section \ref{sub:Distributed-Algorithm-Design} proposes a continuous-time algorithm, and by providing a novel passivity-based perspective of the proposed algorithm, a distributed condition is provided for convergence over time-varying digraphs. In \ref{sub: periodic},  a  periodic communication scheme based on the passivity-based notion is presented, and an event-triggered communication scheme is developed in Section \ref{sub: event}.  The main results are illustrated by an example in Section \ref{sec:4}. Some final remarks and future works are described in Section \ref{sec:5}.}

\section{Preliminaries and Problem Formulation}\label{sec:2}
In this section, we first introduce our notation, some concepts of convex functions and graph theory followed by a passivity-related definition.  Then, the problem to be addressed in this work is formulated. 

\vspace{2mm}
\textbf{Notation} Let $\mathbb{R}$ and $\mathbb{N}$ denote the set of of real numbers and nonnegative integers, respectively. The identity matrix with size $m$ is denoted by $I_{m}$. For symmetric matrices $A$ and $B$, the notation $A\ge B$ ($A>B$) denotes $A-B$ is positive semidefinite (positive definite). $\text{diag}(a_i)$ is the diagonal matrix with $a_i$ being the $i$-th diagonal entry. $\textbf{0}_m$ and $\textbf{1}_m$ denote  all zero and one vectors with size $m\times1$.   For column vectors $v_1,\ldots,v_m$, $\text{col}(v_1,\ldots,v_m)=(v_1^{T},\ldots,v_m^{T})^{T}$. $||\lambda||$ denotes  the Euclidean norm of vector $\lambda$.  
Given a positive semidefinite matrix $A\in\mathbb{R}^{N\times N}$, $\sigma_{min}^{+}(A)$ and $\sigma_{N}(A)$ denote the smallest positive and the largest eigenvalue of $A$, respectively. For a twice differentiable function $f(x)$, its gradient and Hessian are denoted by $\nabla f(x)$ and $\nabla ^{2} f(x)$, respectively. $\text{range}(\nabla f(x))$ denotes the range of the function $\nabla f(x)$. Given a linear mapping $L$, $\text{null}(L)$ denotes the null space of $L$. The Kronecker product is denoted by $\otimes$.

\vspace{2mm}
\textbf{Convex function} A differentiable function $f:\mathbb{R}^{m}\rightarrow \mathbb{R}$ over a convex set $\mathcal{X} \subset  \mathbb{R}^{m}$ is strictly convex if and only if $(\nabla f(x)-\nabla f(y))^T(x-y)>0,\forall x \neq y\in\mathcal{X}$, and it is $\mu$-strongly convex if and only if $(\nabla f(x)-\nabla f(y))^T(x-y)\ge\mu||x-y||,\forall x,y\in\mathcal{X}$, if and only if $f(y)\ge f(x)+\nabla f(x)^T(y-x)+\frac{\mu}{2}||y-x||^2,\forall x,y\in \mathcal{X}$. A function $g:\mathbb{R}^{m}\rightarrow \mathbb{R}^{m}$ over a set $\mathcal{X}$ is $l$-Lipschitz if and only if $||g(x)-g(y)||\le l||x-y||, \forall x,y\in \mathcal{X}$.

\vspace{2mm}
\textbf{Algebraic graph theory } A digraph is a pair $\mathcal{G}=(\mathcal{I},\mathcal{E})$ where $\mathcal{I}={1,\ldots,N}$ is the node set and $\mathcal{E}\subseteq \mathcal{I}\times\mathcal{I}$ is the edge set. An edge $(j,i)\in\mathcal{E}$ means that node $j$ can send information to node $i$, and $i$ is called the out-neighbor of $j$ while $j$ is called the in-neighbor of $i$. A digraph is strongly connected if for every pair of nodes there exists a directed path connecting them. A time-varying graph $\mathcal{G}(t)$ is uniformly jointly strongly connected if there exists a constant $T>0$ such that for any $t_k$, the union $\cup_{t\in [t_k,t_k+T]}\mathcal{G}(t)$ is strongly connected. \textcolor{black}{ A time-varying graph $\mathcal{G}(t)$ is infinitely jointly strongly connected if the union $\cup_{t\in [t,\infty]}\mathcal{G}(t)$ is strongly connected for all $t\ge 0$. Note that  infinitely jointly connected graphs are less restrictive than uniformly jointly strongly connected graphs  as they do not require an upper bound for $T$.}
A weighted digraph is a triple $\mathcal{G}=(\mathcal{I},\mathcal{E},A)$ where $A\in\mathbb{R}^{N\times N}$ is a weighted adjacency matrix defined as $A=[a_{ij}]$ with $a_{ii}=0$, $a_{ij}>0$ if $(i,j)\in \mathcal{E}$ and $a_{ij}=0$, otherwise. The weighted in-degree and  out-degree of node $i$ are $d_{in}^{i}=\sum_{j=1}^{N}a_{ij}$ and $d_{out}^{i}=\sum_{j=1}^{N}a_{ji}$, respectively. A digraph is said to be weight-balanced if $ d_{in}^{i}=d_{out}^{i},\forall i \in \mathcal{I}$. The Laplacian matrix of $\mathcal{G}$ is defined as $L=D_{in}-A$ where $D_{in}=\text{diag}(d_{in}^{i})$.

\vspace{2mm}
\textbf{Input feed-forward passive }
Consider the following nonlinear system:
\[
H:\left\{\begin{array}{rcl}
    \dot{s} &=&F(s,u)  \\
     y &=&Y(s,u), 
\end{array} \right.
\]
where $s\in S\subset \mathbb{R}^{n}$,$u\in U \subset \mathbb{R}^{m}$ and $y\in \mathbb{R}^{m}$ are the state, input and output variables, respectively, and $S,U$ are the state and input spaces, respectively. $F$ and $Y$ are state function and output function.
\begin{definition}(\hspace{1sp}\cite{bao2007process})
System $H$ is \textit{Input feed-forward Passive} (IFP) if there exists a nonnegative real function $V(s):S\rightarrow\mathbb{R}^{+}$, called the storage function, such that for all $t_1\ge t_0 \ge 0$,  initial condition $s_0\in S$ and  $u\in U$,
\begin{equation}\label{eq: def_ifp}
    V(s(t_1))-V(s(t_0))\le \int_{t_0}^{t_1}u^{T}y-\nu u^{T}udt
\end{equation}
for some $\nu\in\mathbb{R}$, denoted as IFP($\nu$).  
\end{definition}

If the storage function $V(s)$ is differentiable, the inequality  \eqref{eq: def_ifp} is equivalent to
\begin{equation}
    \dot{V}(s)\le u^{T}y-\nu u^{T}u.
\end{equation}

As it can be seen from the above definition, a positive value of $\nu$ means that the system has an excess of passivity while a negative value of $\nu$ means  the system lacks  passivity. The index $\nu$ can be taken as a measurement to quantify how passive a dynamic system is.  This concept  will play a crucial role in the subsequent results.

\vspace{2mm}
\textbf{Problem formulation}  Each node $i$ has a local cost function $f_{i}(x_{i}):\mathbb{R}^{m}\rightarrow\mathbb{R}$ where $x_{i}\in\mathbb{R}^{m}$ is the local decision variable. The sum of $f_i(x_i)$ is considered as the global cost function. We make the following assumptions.
\vspace{2mm}
\begin{assum}\label{assum:1}
Each $f_{i},i\in\mathcal{I}$ is twice differentiable with $\nabla^{2}f_{i}(x_i)>0$ and its gradient $\nabla f_{i}(x_i)$ is $l_{i}$-Lipschitz.
\end{assum} 
Under Assumption \ref{assum:1}, $f_i$ is strictly convex and 
\begin{equation}
||\nabla f_{i}(x_i)-\nabla f_{i}(y_i)||\le l_{i}||x_i-y_i||.\label{eq:assumption_1}
\end{equation}
Thus, its Hessian satisfies 
\begin{equation}
0<\nabla^{2}f_{i}(x_i)\le l_{i}I,\forall i\in\mathcal{I}.\label{eq:assumtion_3}
\end{equation}
\begin{assum}\label{assum:3}
The time-varying communication graph $\mathcal{G}(t)$ is weight-balanced and \textcolor{black}{infinitely} jointly strongly connected.
\end{assum} 

The objective is to design a continuous-time distributed algorithm such that the following problem
\begin{equation}
\begin{array}{rl}
 \underset{x_{1},\ldots,x_{N}}{\text{min}}  & \displaystyle \sum_{i=1}^{N}f_{i}(x_{i}) \\
  \text{s.t. } & \displaystyle  \sum_{i=1}^{N}x_{i}=\sum_{i=1}^{N}d_{i}
\end{array}\label{eq:resource allocation problem}
\end{equation}
is solved by each node using only its own information and exchanged information from its neighbors under discrete-time communication. In fact, this problem can be used to formulate many practical applications such as network utility maximization and economic dispatch in power systems.

Let us denote $x=\text{col}(x_{1},\ldots,x_{N})$. It can be observed that 
problem \eqref{eq:resource allocation problem} is feasible and has
a unique optimal point $x^{*}.$

\section{Main Results}\label{sec:3}
 
\subsection{\label{sub:The-Lagrange-dual}The Lagrange dual problem}
In this subsection, we show that the resource allocation problem \eqref{eq:resource allocation problem} can be equivalently converted into a general distributed convex optimization. 

Let us define a set of new variable $\lambda_i\in\mathbb{R}^{m},i\in\mathcal{I}$, and denote the set of  $\text{range}(\nabla f_i)$ as $\Lambda_i$. It can be derived from \cite{minty1964monotonicity} that $\Lambda_i$ is a convex set. Under Assumption \ref{assum:1}, we have that the inverse function of $\nabla f_i(\cdot)$ exists and is differentiable,  denoted as $h_i(\cdot)$  \textcolor{black}{\footnote{If the analytic form
of the inverse function $h_i(\cdot)$ can not be obtained, one can replace $h_i(\cdot)$ by ${\text{argmin}_{x_i}} \{f_i(x_i)-\lambda_i^{T}x_i \}$  in the algorithms that will be proposed later. This replacement does not affect our analysis. }}, and further  define
\begin{equation}
g_{i}(\lambda_i)\triangleq f_{i}(h_{i}(\lambda_i))+\lambda_i^{T}\left(d_{i}-h_{i}(\lambda_i)\right)\label{eq:g_i}
\end{equation}
when $\lambda_i\in\Lambda_i$.
\vspace{2mm}
\begin{lem}\label{lem:0}
Problem \eqref{eq:resource allocation problem} can be equivalently solved by the following convex optimization
\begin{equation}
\begin{array}{cl}
  \underset{\lambda_{i}\in \Lambda_i,\forall i\in\mathcal{I}}{\text{min}} & J(\lambda)=\displaystyle\sum_{i=1}^{N}J_{i}(\lambda_{i})\\
 \text{s.t.} & \;\lambda_{i}=\lambda_{j},\forall i,j\in\mathcal{I}
\end{array}\label{eq:distributed convex optimization}
\end{equation}
with $J_{i}(\lambda_{i})=-g_{i}(\lambda_{i})$ and $\nabla J_{i}(\lambda_{i})=h_{i}(\lambda_{i})-d_{i}$. Moreover,  $J_{i}(\lambda_{i})$ is twice differentiable and
$\frac{1}{l_{i}}$-strongly convex in the domain $\Lambda_i$, i.e., $\frac{1}{l_i}\le\nabla^{2}J_i(\lambda_i),\forall \lambda_i\in\Lambda_i$. 
\end{lem}

\textbf{Proof.}
This result can be obtained via the duality \cite{bertsekas1996neuro}. 

\endprf
\vspace{2mm}

Due to the strong duality, the primal optimal solution $x^{*}$ is a minimizer
of $\mathcal{L}(x,\lambda^{*})$ which is defined as
\begin{equation}
\mathcal{L}(x,\lambda^{*})=\sum_{i=1}^{N}f_{i}(x_{i})+{\lambda^*} ^{T}\left(\sum_{i=1}^{N}d_{i}-\sum_{i=1}^{N}x_{i}\right)\label{eq:lagrangian}
\end{equation}

This fact enables us to recover the primal solution $x^{*}$ from
the dual optimal solution $\lambda^{*}$. Specifically, since $f_{i}$
is strictly convex, the function $\mathcal{L}(x,\lambda^{*})$ is
strictly convex in $x$, and therefore has a unique minimizer which
is identical to $x^{*}$. Moreover, since $\mathcal{L}(x,\lambda^{*})$
is separable according to \eqref{eq:lagrangian}, we can recover $x_{i}^{*}$
from  $x_{i}^{*}=h_{i}(\lambda^{*})$.

Based on Lemma \ref{lem:0}, we then aim to design an continuous-time algorithm for problem \eqref{eq:distributed convex optimization}. For simplicity, we will abuse the notation by using $\lambda=\text{col}(\lambda_1,\ldots,\lambda_N)$ hereafter.

\subsection{\label{sub:Distributed-Algorithm-Design}IFP-based Distributed Algorithm Design
}

For $i\in\mathcal{I}$
and with constant scalars $\alpha,\beta>0$, let us consider the following continuous-time algorithm
\begin{equation}
\begin{array}{rcl}
\dot{\lambda_{i}} & = & -\alpha(h_{i}(\lambda_{i})-d_{i})-\gamma_{i}\\
\dot{\gamma}_{i} & = & -u_{i}\\
u_{i} & = & \beta\sum_{j=1}^{N}a_{ij}(t)(\lambda_{j}-\lambda_{i})
\end{array}\label{eq:algorithm}
\end{equation}
where $\lambda_{i},\gamma_{i}\in\mathbb{R}^{m}$ are the local states
variables and $u_{i}\in\mathbb{R}^{m}$ is the local input. $\alpha>0$
is a predefined constant and $\beta>0$ is the coupling
gain to be designed. $A(t)=[a_{ij}(t)]_{N\times N}$ is the adjacency matrix
of the graph $\mathcal{G}(t)$. 

Let $\gamma=\text{col}(\gamma_{1},\ldots\gamma_{N})$,
$d=\text{col}(d_{1},\ldots,d_{N})$ and $h(\lambda)=\text{col}(h_{1}(\lambda_{1}),\ldots,h_{N}(\lambda_{N}))$.
The algorithm \eqref{eq:algorithm} can be rewritten in a compact
form as 
\begin{equation}
\begin{array}{rcl}
\dot{\lambda} & = & -\alpha\left(h(\lambda)-d\right)-\gamma\\
\dot{\gamma} & = & \beta\mathbf{L}(t)\lambda
\end{array}\label{eq:dynamic_compact form}
\end{equation}
where $\mathbf{L}(t)=L(t)\otimes I_{m}$ with $L(t)$ being the Laplacian matrix
of the graph $\mathcal{G}(t)$. 

The above continuous-time algorithm is a simplification of the one proposed in \cite{kia2015distributed} which is motivated by the feedback control consideration. Specifically, each node evolves in the direction of gradient decent while trying to reach an agreement with its neighbors. To correct the error between the local gradient and the consensus with neighbors, the integral feedback of $u_i$ representing the node disagreements is exploited.  An important reason for using such an algorithm is that it enables us to provide a passivity-based perspective for the individual algorithmic dynamics later.

 In the rest of this work, we assume that   $\lambda_i(0)\in\Lambda_i$ for all $i\in\mathcal{I}$. This can be trivially satisfied by letting $\lambda_i(0)=\nabla f_i(x_i(0))$.   

In the following, we will first show in Lemma \ref{lem:1} that the optimal solution of \eqref{eq:distributed convex optimization} coincides with the equilibrium  point of algorithm \eqref{eq:algorithm}. Then we provide a passivity-based perspective for the error dynamics in each individual node in Theorem \ref{thm:IFP}, based on which the convergence of algorithm \eqref{eq:algorithm} is shown in Theorem \ref{thm:convergence}. 

\vspace{2mm}
\begin{lem}
\label{lem:1}Under Assumptions \ref{assum:1} and \ref{assum:3}, the equilibrium point $(\lambda^{*},\gamma^{*})$
of the system \eqref{eq:algorithm} with the initial condition $\sum_{i=1}^{N}\gamma_{i}(0)=\mathbf{0}$
is unique and $\lambda^{*}$ is the optimal solution of problem \eqref{eq:distributed convex optimization}.\end{lem}
\textbf{Proof.}
Suppose $(\lambda^{*},\gamma^{*})$ is the equilibrium of system \eqref{eq:algorithm}
and $\sum_{i=1}^{N}\gamma_{i}(0)=\mathbf{0}$. It follows that 
\begin{equation}
\begin{array}{rcl}
\dot{\lambda^{*}} & = & -\alpha\left(h(\lambda^{*})-d\right)-\gamma^{*}=\mathbf{0}\\
\dot{\gamma^{*}} & = & \beta\mathbf{L}(t)\lambda^{*}=\mathbf{0}
\end{array}.\label{eq: equilibrium}
\end{equation}
Since $\left(1_{N}\otimes I_{m}\right)^{T}\mathbf{L}(t)=\mathbf{0}_{Nm}^{T}$,
we have $\left(1_{N}\otimes I_{m}\right)^{T}\dot{\gamma}=\beta\left(1_{N}\otimes I_{m}\right)^{T}\mathbf{L}(t)\lambda=\mathbf{0}$,
which gives $\sum_{i=1}^{N}\dot{\gamma}_{i}=\mathbf{0}$. Hence, it can be
observed that $\sum_{i=1}^{N}\gamma_{i}(t)=\sum_{i=1}^{N}\gamma_{i}(0)=\mathbf{0}_{m}$
for any $t\ge0$. Next, let us multiply $\left(1_{N}\otimes I_{m}\right)^{T}$
from the left of the $\dot{\lambda}^{*}$, and obtain that 
\[
\begin{array}{ll}
& \left(1_{N}\otimes I_{m}\right)^{T}\dot{\lambda^{*}}\\=&-\alpha\left(1_{N}\otimes I_{m}\right)^{T}\left(h(\lambda^{*})-d\right)-\sum_{i=1}^{N}\gamma_{i}^{*}
 =\mathbf{0},
\end{array}
\]
 which indicates that 
\[
\nabla J(\lambda^{*})=\sum_{i=1}^{N}\nabla J_{i}(\lambda_{i}^{*})=\sum_{i=1}^{N}\left(h_{i}(\lambda_{i}^{*})-d_{i}\right)=\mathbf{0}.
\]
Moreover, since the graph $\mathcal{G}(t)$ is \textcolor{black}{infinitely} jointly strongly connected, 
$\dot{\gamma^{*}}=\beta\mathbf{L}(t)\lambda^{*}\equiv \mathbf{0}$ implies that $\lambda_{1}^{*}=\ldots=\lambda_{N}^{*}$.
Under Assumption \ref{assum:1},  problem \eqref{eq:distributed convex optimization}
has a unique solution, which coincides with $\lambda^{*}$ based on
the optimality condition \cite{ruszczynski2006nonlinear}. 

\endprf

\vspace{2mm}
Before proceeding to show in Theorem \ref{thm:convergence} that the algorithm converges, let us investigate the IFP property of the error dynamics in each individual node.  Denote $\Delta\lambda_{i}=\lambda_{i}-\lambda_{i}^{*}$ and
$\Delta\gamma_{i}=\gamma_{i}-\gamma_{i}^{*}$. Comparing \eqref{eq:algorithm} and
\eqref{eq: equilibrium} yields the individual error system shown
as 
\begin{equation}
\Psi_{i}:\left\{ \begin{array}{rcl}
\Delta\dot{\lambda}_{i} & = & -\alpha\left(h_{i}(\lambda_{i})-h_{i}(\lambda_i^{*})\right)-\Delta\gamma_{i}\\
\Delta\dot{\gamma}_{i} & = & -u_{i}\\
u_{i} & = & \beta\sum_{j=1}^{N}a_{ij}(t)(\Delta\lambda_{j}-\Delta\lambda_{i}).
\end{array}\right.\label{eq: Psi}
\end{equation}
By taking $u_{i}$ and $\Delta\lambda_{i}$ as the input and output
of the error system $\Psi_{i}$, the following theorem shows that
each error system $\Psi_{i}$ is IFP with the proof  provided in Appendix.

\vspace{2mm}
\begin{thm}
\label{thm:IFP}Suppose Assumption \ref{assum:1} holds. Then, the system $\text{\ensuremath{\Psi}}_{i}$
is IFP($\nu_{i}$) from $u_{i}$ to $\Delta\lambda_{i}$ with $\nu_{i}\ge-\frac{l_{i}^{2}}{\alpha^{2}}$.\end{thm}
\vspace{2mm}

\begin{rem}
It is shown in the above theorem that for the nonlinear system \eqref{eq: Psi}
resulting from general strongly convex objective  function $J_{i}(\lambda_{i})$
is IFP from $u_{i}$ to $\Delta\lambda_{i}$. Moreover, the IFP index
is lower bounded by $-\frac{l_{i}^{2}} {\alpha^{2}}$, which
means that the system \eqref{eq: Psi} can have the IFP index arbitrarily
close to 0 (i.e, passivity) if the coefficient $\alpha$ can take
arbitrarily large value. However, it might be impractical to choose
an infinitely large $\alpha$ due to the potential numerical error
or larger computing costs when solving the ordinary differential equation \eqref{eq:dynamic_compact form}
numerically. In view of this, in order to achieve larger IFP index,
we can choose $\alpha$ as the largest positive number allowed by
the error tolerance error level of the available computing platform. 
It is worth mentioning that similar algorithm with \eqref{eq:algorithm} has been shown in \cite{kia2015distributed}. The contribution of Theorem \ref{thm:IFP} is to provide a novel passivity-based perspective of the proposed algorithm, and this perspective will lead to fruitful results in the remainder of this section.
\end{rem}

The next theorem provides a condition to design the coupling gain
$\beta$ under which the algorithm \eqref{eq:algorithm} will converge
to the optimal solution of  problem \eqref{eq:distributed convex optimization}.

\vspace{2mm}
\begin{thm}
\label{thm:convergence}Under Assumptions \ref{assum:1} and \ref{assum:3}, suppose the coupling gain
$\beta$ satisfies 
\begin{equation}
0<\beta<\frac{\alpha^{2}\sigma_{min}^{+}\left(L(t)+L(t)^{T}\right)}{2\sigma_{N}\left(L(t)^{T}\text{diag}\left(l_{i}^{2}\right)L(t)\right)},\label{eq:beta_condition}
\end{equation}
where $\sigma_{min}^{+}$ and $\sigma_N$ are the smallest positive and the largest eigenvalue respectively. Then under algorithm \eqref{eq:algorithm}, for all $i\in\mathcal{I}$, the set $\Lambda_i$ is a positively invariant set of $\lambda_i$, and  the algorithm \eqref{eq:algorithm} with any initial condition
with $\sum_{i=1}^{N}\gamma_{i}(0)=\mathbf{0}$ will converge to the optimal
solution of \eqref{eq:distributed convex optimization}. \end{thm}
\textbf{Proof.} The proof is  stated in  Appendix.\endprf
\vspace{2mm}

\begin{rem}
Lemma \ref{lem:1} states that the equilibrium point of the continuous-time
algorithm \eqref{eq:algorithm} under the initial constraint $\sum_{i=1}^{N}\gamma(0)=\mathbf{0}$
is identical to the optimal solution of the distributed optimization
problem \eqref{eq:distributed convex optimization} while Theorem
\ref{thm:convergence} states that the algorithm \eqref{eq:algorithm}
will converge to such an equilibrium point if the coefficients 
$\alpha$ and $\beta$ are chosen to satisfy \eqref{eq:beta_condition}.
As discussed in Section \ref{sub:The-Lagrange-dual}, the optimal solution
$x_{i}^{*}$ of the original resource allocation problem \eqref{eq:resource allocation problem}
can be recovered from $x_{i}^{*}=h_{i}(\lambda^{*})$.
In this view, the distributed algorithm in \eqref{eq:algorithm} utilizes
only local interaction with exchanging $\lambda_{i}$ instead of the
real decision variable $x_{i}$ to achieve the optimal collective
goal. 
\end{rem}
\vspace{2mm}
It should be mentioned that the condition proposed in Theorem \ref{thm:convergence} maybe difficult to be examined in a time-varying graph. Nevertheless, the following distributed condition can be obtained based on Theorem \ref{thm:convergence}.
\vspace{2mm}
\begin{cor}\label{cor: conditions}
Under Assumptions \ref{assum:1} and \ref{assum:3}, the algorithm \eqref{eq:algorithm} with any initial condition with $\sum_{i=1}^{N}\gamma_i(0)=\mathbf{0}$ will converge to the optimal solution of \eqref{eq:distributed convex optimization} if the coupling gain $\beta > 0$ satisfies :
   \begin{equation}
        \beta \frac{l_{i}^{2}}{\alpha^{2}}d_{in}^{i}(t) <\frac{1}{2}, \forall i\in \mathcal{I},\forall t>0\label{eq: individual condition}
    \end{equation}
where $d_{in}^{i}(t)$ denotes the in-degree of the $i$-th node. 
\end{cor}
\textbf{Proof.} The proof is stated in  Appendix.\endprf

\vspace{2mm}
\begin{rem}\label{re:design}(Design of parameter $\beta$)
In order to implement the algorithm \eqref{eq:algorithm},  the parameter $\beta$ needs to be designed. The condition proposed in the above corollary provides a distributed strategy to design $\beta$. A heuristic solution is to let each node compute the maximum $\beta$ according to \eqref{eq: individual condition} and search the minimum of $\beta$ among them by communicating among neighboring nodes. Repeat this procedure when a smaller $\beta$ is updated (a larger $d_{in}^{i}(t)$ is detected) at any node due to the graph variation.   However, this has to be done in an off-line manner. A possible more easy way is to let the upper bound of $\beta$ be $\frac{\alpha ^2}{2 \text{max}_i \left\lbrace l_i \right\rbrace (N-1)}$ when $a_{ij} \le 1$, $\forall ~i,~j$. 
\end{rem}

\subsection{ Periodic Discrete-time Communication} \label{sub: periodic}

Continuous-time communication among the nodes is required in the distributed algorithm proposed in Section \ref{sub:Distributed-Algorithm-Design}
whereas a digital network with limited channel capacity generally allows communication only at discrete instants. Moreover, the communication cost is far larger than the computation cost in real applications like sensor networks \cite{wan2009event}. 
To separate the communication and the computation, we will investigate in this subsection the distributed algorithm design under periodic discrete-time communication by exploiting the IFP property stated in Theorem \ref{thm:IFP}.

By considering a  sampling based scheme, we proceed to investigate the convergence of algorithm \eqref{eq:algorithm} with periodic communication. 
\begin{figure}[tbh]
\centering
\includegraphics[width = 1\linewidth]{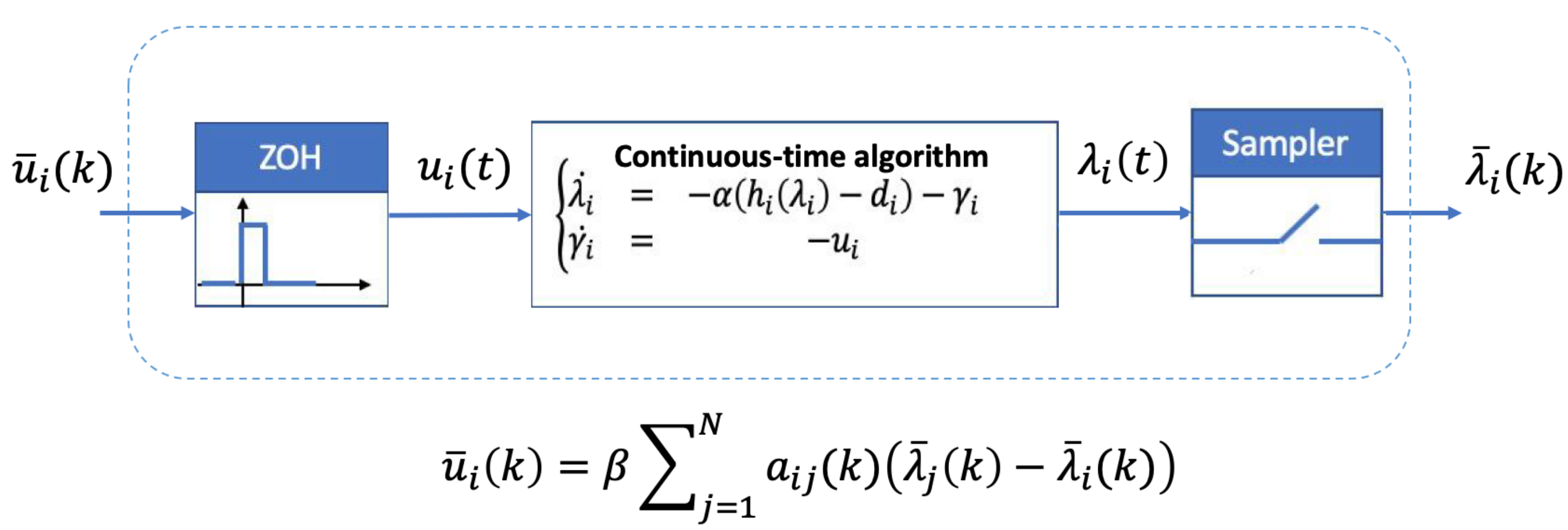}\center\caption{Sampled continuous distributed algorithm. }\label{fig:1}
\end{figure}

As depicted in Figure \ref{fig:1}, let us consider
the algorithm with sampling at each output of individual node, 
\begin{equation}
\begin{array}{rcl}
\dot{\lambda_{i}} & = & -\alpha(h_{i}(\lambda_{i})-d_{i})-\gamma_{i}\\
\dot{\gamma}_{i} & = & -u_{i}\\
\bar{u}_{i}\textcolor{black}{(k)} & = & \beta\sum_{j=1}^{N}a_{ij}(k)(\bar{\lambda}_{j}\textcolor{black}{(k)}-\bar{\lambda}_{i}\textcolor{black}{(k)})
\end{array}\label{eq:sampled algorithm}
\end{equation}
\textcolor{black}{where $a_{ij}(k)$ denotes $a_{ij}(t)$ at the $k$-th sampling instant, }the output $\bar{\lambda}_{i}$ is obtained by sampling
the continuous-time output $\lambda_{i}$, while the input $u_{i}$ depending
on the sampled $\bar{\lambda}_{i},\forall i\in\mathcal{I}_{i}$ is
applied to the continuous-time system through a zero order holder. In particular,
let the sampling period be denoted as $T_{s}$, and then for all $k\in\mathbb{N}$,
\begin{equation}
\begin{array}{l}
\bar{\lambda}_{i}(k)  =  \lambda_{i}(kT_{s}),\\
u_{i}(t) = \bar{u}_{i}(k),\forall t\in[kT_{s},(k+1)T_{s}).
\end{array}\label{eq:sampling}
\end{equation}

Since the communication is carried out in periodic discrete-time instants, we need to make the following additional assumption for the graph. Denote the time sequence $k=\{0,T_s,2T_s,\ldots\}$.
\vspace{2mm}
\begin{assum}\label{assum 2}
The time-varying graph $\mathcal{G}(k)$ is balanced and \textcolor{black}{infinitely} jointly strongly connected, i.e., \textcolor{black}{there exists an unbounded time sequence $k,k+1,k+2,\ldots$  such that $\mathcal{G}(k)\cup \mathcal{G}(k+1)\cup \mathcal{G}(k+2)\cup\cdots$ }is strongly connected for any $k\in \mathbb{N}$.
\end{assum} 
\vspace{1mm}
With $\Delta\bar{\lambda}_{i}=\bar{\lambda}_{i}-\lambda_{i}^{*}$
where $\lambda_{i}^{*}$ is defined in \eqref{eq: equilibrium}, the
error dynamics of subsystem $i$ is 
\begin{equation}
\bar{\Psi}_{i}:\left\{ \begin{array}{rcl}
\Delta\dot{\lambda}_{i} & = & -\alpha\left(h_{i}(\lambda_{i})-h_{i}(\lambda^{*})\right)-\Delta\gamma_{i}\\
\Delta\dot{\gamma}_{i} & = & -u_{i}\\
\bar{u}_{i} & = & \beta\sum_{j=1}^{N}a_{ij}(\Delta\bar{\lambda}_{j}-\Delta\bar{\lambda}_{i}).
\end{array}\right.\label{eq:sampling Psi}
\end{equation}

In the following, we first analyze and approximate the bound of the sampling error $\Delta \lambda_i-\Delta\bar{\lambda}_i$ with respect to the input $\bar{u}_i$ in Lemma \ref{lem:l2gain} and \ref{lem:sampled error}. Based on these results, Theorem \ref{thm:ifp index_sampled system} characterizes the passivity degradation over sampling of the error dynamics at each node, and the convergence of the algorithm \eqref{eq:sampled algorithm} is stated in Corollary \ref{cor: sample cond}.

For notational simplicity, let us denote $z_{i}=\Delta\dot{\lambda}_{i}$. 
\vspace{2mm}
\begin{lem}
\label{lem:l2gain}Suppose Assumption \ref{assum:1} holds. 
Then, under the dynamics $\bar{\Psi}_{i}$, it holds that for all $u_{i}\in\mathbb{R}^{m}$,
\begin{equation}
\frac{l_i}{\alpha}\cdot \frac{d||z_{i}||^{2}}{dt}\le \frac{l_{i}^{2}}{\alpha^2}||u_{i}||^{2}-||z_{i}||^{2}.\label{eq:l2gain}
\end{equation}
\end{lem}
\textbf{Proof.}
The derivative of $z_{i}$ yields that 
\[
\dot{z}_{i}=-\alpha\frac{\partial h_{i}(\lambda_{i})}{\partial\lambda_{i}}z_{i}-\Delta\dot{\gamma}_{i}=-\alpha\frac{\partial h_{i}(\lambda_{i})}{\partial\lambda_{i}}z_{i}+u_{i}
\]
and it leads to 
\[
\frac{l_i}{\alpha} \cdot \frac{d||z_{i}||^{2}}{dt}=2\frac{l_i}{\alpha} z_{i}^{T}\left(-\alpha\frac{\partial h_{i}(\lambda_{i})}{\partial\lambda_{i}}z_{i}+u_{i}\right).
\]
We can also observe that
\[
\begin{pmatrix}\frac{2\alpha }{l_{i}}\frac{l_i}{\alpha}-1 & -\frac{l_i}{\alpha}\\
-\frac{l_i}{\alpha} & \frac{l_{i}^{2}}{\alpha ^{2}}
\end{pmatrix}\ge0,
\]
which follows that 
\[
\left(\begin{array}{c}
z_{i}\\
u_{i}
\end{array}\right)^{T}\left(\begin{pmatrix}\frac{2\alpha }{l_{i}}\frac{l_i}{\alpha}-1 & -\frac{l_i}{\alpha}\\
-\frac{l_i}{\alpha}& \frac{l_{i}^{2}}{\alpha ^{2}}
\end{pmatrix}\otimes I_{m}\right)\left(\begin{array}{c}
z_{i}\\
u_{i}
\end{array}\right)\ge0,\forall z_{i},u_{i}.
\]
Since $\frac{1}{l_{i}}I_m\le\frac{\partial h_{i}(\lambda_{i})}{\partial\lambda_{i}}$ under Assumption \ref{assum:1}, we further obtain that for all $z_{i},u_{i}\in\mathbb{R}^{m}$ 
\[
\left(\begin{array}{c}
z_{i}\\
u_{i}
\end{array}\right)^{T}\begin{pmatrix}2\frac{l_{i}}{\alpha } \left(\alpha\frac{\partial h_{i}(\lambda_{i})}{\partial\lambda_{i}}\right)-I_m & - \frac{l_{i}}{\alpha }I_m  \\
-\frac{l_{i}}{\alpha } I_m & \frac{l_{i}^{2}}{\alpha ^{2}}I_m
\end{pmatrix}\left(\begin{array}{c}
z_{i}\\
u_{i}
\end{array}\right)\ge0,
\]
which is equivalent to $\frac{l_{i}}{\alpha }\frac{d||z_{i}||^{2}}{dt}\le \frac{l_{i}^{2}}{\alpha ^{2}} ||u_{i}||^{2}-||z_{i}||^{2}$. 
\endprf

\vspace{2mm}
From the above lemma, it can be seen by the integration of \eqref{eq:l2gain} over $t\in[kT_{s},(k+1)T_{s}]$
that 
\begin{equation}
\begin{array}{l}
\frac{l_{i}}{\alpha}||z_{i}((k+1)T_{s})||^{2}-\frac{l_{i}}{\alpha }||z_{i}(kT_{s})||^{2}\\
\le \frac{l_{i}^{2}}{\alpha ^{2}}  \int_{kT_{s}}^{(k+1)T_{s}}||u_{i}(t)||^{2}dt-\int_{kT_{s}}^{(k+1)T_{s}}||z_{i}(t)||^{2}dt.
\end{array}\label{eq:l2gain of derivative}
\end{equation}
It can be seen from the form of \eqref{eq:l2gain} or \eqref{eq:l2gain of derivative}
that $\frac{l_{i}^{2}}{\alpha ^{2}} $ provides the upper bound of the $\mathcal{L}_{2}$ gain for
the mapping $u_{i}\rightarrow z_{i}$ since the specific form of storage
function, $\frac{l_{i}}{\alpha} ||z_{i}||^{2}$, is considered. 

\vspace{2mm}
\begin{lem}
\label{lem:sampled error}Under Assumption \ref{assum:1}, for all $k\in\mathbb{N},$
the following inequality holds
\begin{equation}
\begin{array}{l}
\int_{kT_{s}}^{(k+1)T_{s}}||\Delta\lambda_{i}(t)-\Delta\bar{\lambda}_{i}(k)||^{2}dt\le T_{s}^{2}\cdot\\
\left(T_{s}\frac{l_{i}^{2}}{\alpha ^{2}}||\bar{u}_{i}(k)||^{2}+\frac{l_{i}}{\alpha}\left(||z_{i}(kT_{s})||^{2}-||z_{i}((k+1)T_{s})||^{2}\right)\right).
\end{array}\label{eq:sampling error lemma}
\end{equation}
\end{lem}

\textbf{Proof.}
First, let us observe that for all $t\in[kT_{s},(k+1)T_{s}),\forall k\in\mathbb{N}$,
\begin{eqnarray}
\left| \left|\int_{kT_{s}}^{t}\Delta\dot{\lambda}_{i}(s)ds\right| \right| ^{2} & \le & \left|\left| \int_{kT_{s}}^{(k+1)T_{s}}\left|\left| \Delta\dot{\lambda}_{i}(s)\right|\right| ds\right|\right| ^{2}\nonumber \\
 & \le & T_{s}\int_{kT_{s}}^{(k+1)T_{s}}\left|\left| \Delta\dot{\lambda}_{i}(s)\right| \right|^{2}ds\label{eq:sampling error proof1}
\end{eqnarray}
where the second inequality holds based on Cauchy-Schwarz inequality.

Next, it follows from \eqref{eq:l2gain of derivative} and \eqref{eq:sampling error proof1}
that 
\[
\begin{array}{ll}
 & \int_{kT_{s}}^{(k+1)T_{s}}||\Delta\lambda_{i}(t)-\Delta\bar{\lambda}_{i}(k)||^{2}dt\\
= & \int_{kT_{s}}^{(k+1)T_{s}}||\int_{kT_{s}}^{t}\Delta\dot{\lambda}_{i}(s)ds||^{2}dt\\
\le & \int_{kT_{s}}^{(k+1)T_{s}}\left(T_{s}\int_{kT_{s}}^{(k+1)T_{s}}\left| \left|\Delta\dot{\lambda}_{i}(s)\right|\right| ^{2}ds\right)dt\\
= & T_{s}^{2}\int_{kT_{s}}^{(k+1)T_{s}}\left|\left| \Delta\dot{\lambda}_{i}(s)\right| \right|^{2}ds\\
\le & T_{s}^{2}\frac{l_{i}^{2}}{\alpha ^{2}} \int_{kT_{s}}^{(k+1)T_{s}}||u_{i}(s)||^{2}ds+T_{s}^{2}\frac{l_{i}}{\alpha}\cdot\\
 & \left(||z_{i}(kT_{s})||^{2}-||z_{i}((k+1)T_{s})||^{2}\right).
\end{array}
\]
Based on the relationship between $u_{i}(t)$ and $\bar{u}_{i}(k)$
shown in \eqref{eq:sampling}, the inequality \eqref{eq:sampling error lemma}
can be therefore obtained.\endprf

\vspace{2mm}
\begin{thm}
\label{thm:ifp index_sampled system}Under Assumption \ref{assum:1}, the sampled
system $\bar{\Psi}_{i}$ is IFP$(\bar{\nu}_{i})$ from $\bar{u}_{i}$
to $\Delta\bar{\lambda}_{i}$ with $\bar{\nu}_{i}\ge-\left(\frac{l_{i}^{2}}{\alpha^{2}}+T_{s}\frac{l_{i}}{\alpha}\right)$
where $T_{s}$ is the sampling period.\end{thm}
\textbf{Proof.} The proof is stated in  Appendix.\endprf

\vspace{2mm}
Theorem \ref{thm:ifp index_sampled system} shows that the lower bound of the IFP index, $\nu$, decreases from  $-\frac{l_{i}^{2}}{\alpha^{2}}$ to $-\frac{l_{i}^{2}}{\alpha^{2}}-T_{s}\frac{l_{i}}{\alpha}$ over the sampling. This passivity ''degradation'' is caused by sampling error, which depends on the sampling period $T_s$. Based on this new IFP index bound, a revised distributed condition for convergence of the algorithm \eqref{eq:sampled algorithm} is provided as follows.

\vspace{2mm}
\begin{cor}\label{cor: sample cond}
Under Assumptions \ref{assum:1} and  \ref{assum 2}, the algorithm \eqref{eq:sampled algorithm} under periodic communication
with any initial condition with $\sum_{i=1}^{N}\gamma_{i}(0)=\mathbf{0}$ will
converge to the optimal solution of \eqref{eq:distributed convex optimization} if the following condition is satisfied for all $t\ge 0$:
\begin{equation}\label{eq:beta_cond_sampling}
\beta \left(\frac{l_{i}^{2}}{\alpha^{2}}+T_{s}\frac{l_{i}}{\alpha}\right)d_{in}^{i}(t)<\frac{1}{2},\forall i\in \mathcal{I}.
\end{equation}
 \end{cor}

 \textbf{Proof.}
 This condition can be derived based on  similar argument in the proofs of Theorem \ref{thm:convergence}
 and Corollary \ref{cor: conditions}, and the discrete-time  LaSalle's invariance principle \cite{mei2017lasalle}.\endprf
 
\vspace{2mm}
 As shown in the above corollary, when $\alpha$ and $\beta$ are fixed and satisfy the condition \eqref{eq: individual condition}, \textcolor{black}{there always exists a constant $T_s > 0$ satisfying \eqref{eq:beta_cond_sampling}}. Indeed, with fixed $\alpha$ and $\beta$, the sampling period $T_s$ can also be determined in a distributed way by a similar heuristic solution described in Remark  \ref{re:design}. \textcolor{black}{It is worth noting that a sufficiently large sampling period $T_s$ is acceptable provided $\alpha$ is large enough and coupling gain $\beta$ is small enough.}

 \subsection{\textcolor{black}{Distributed Event-Driven Communication}\label{sub: event}}
\textcolor{black}{Build on the sampled-based  framework in the preceding subsection, we further consider   an event-driven communication strategy.  
\begin{figure}[tbh]
\centering
\includegraphics[width = 1\linewidth]{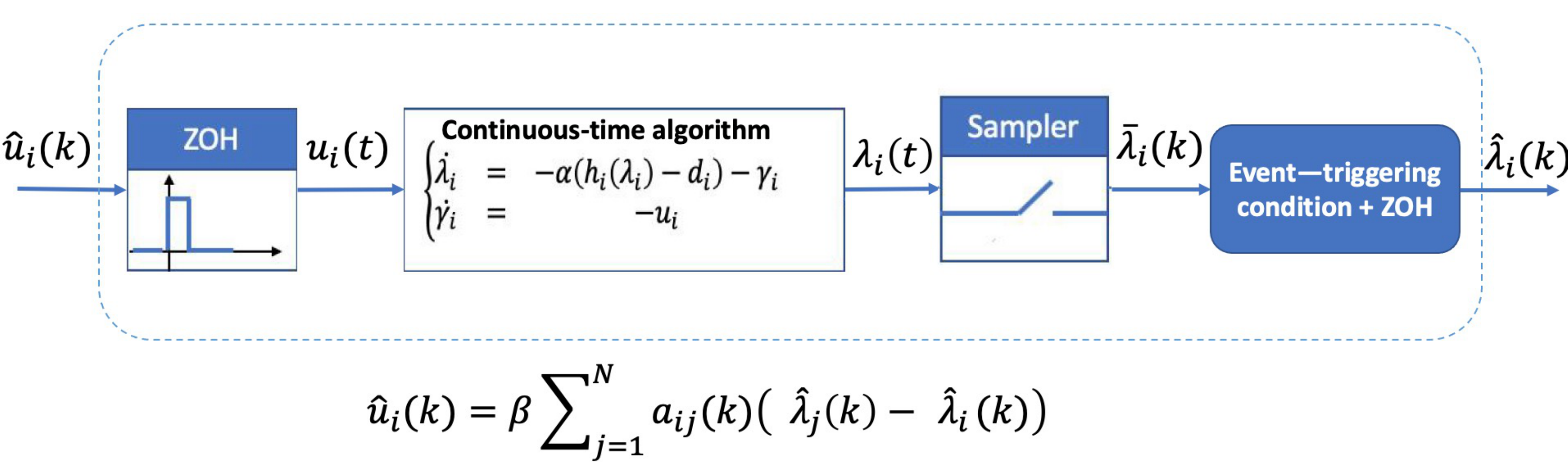}\center \caption{ Continuous distributed algorithm with sampled-based event-triggered communication.}\label{fig:2}
\end{figure}
Reconsider the algorithm as shown in \eqref{eq:sampled algorithm} by incorporating an event-driven communication mechanism depicted by Figure \ref{fig:2}, i.e.,  
\begin{equation}
\begin{array}{rcl}
\dot{\lambda_{i}} & = & -\alpha(h_{i}(\lambda_{i})-d_{i})-\gamma_{i}\\
\dot{\gamma}_{i} & = & -u_{i}\\
\hat{u}_{i}\textcolor{black}{(k)} & = & \beta\sum_{j=1}^{N}a_{ij}(k)(\hat{\lambda}_{j}\textcolor{black}{(k)}-\hat{\lambda}_{i}\textcolor{black}{(k)})
\end{array}\label{eq:algorithm-1}
\end{equation}
where $\hat{\lambda}_{i}\textcolor{black}{(k)},\ i\in\mathcal{I}$ denotes the last known
state of node $i$ that has been transmitted to its neighbors at the time $kT_s$. Similar to \eqref{eq:sampling}, we set
\begin{equation}
\begin{array}{l}
\bar{\lambda}_{i}(k)  =  \lambda_{i}(kT_{s}),\\
u_{i}(t) = \hat{u}_{i}(k),\forall t\in[kT_{s},(k+1)T_{s}).
\end{array}
\end{equation}}

\textcolor{black}{The following theorem presents a triggering condition for each node to update its output while the convergence to the global  optimal solution is ensured. }

\vspace{2mm}
\textcolor{black}{
\begin{thm}\label{thm:event}
Under Assumptions \ref{assum:1} and  \ref{assum 2}, consider the algorithm \eqref{eq:algorithm-1}.
If  $\alpha,\beta$ are designed such that \eqref{eq:beta_cond_sampling} holds,
 and the triggering instant for node $i,i\in \mathcal{I}$ to transmit its
current information of $\lambda_{i}$ is chosen whenever the following
condition is satisfied 
\begin{equation}
\begin{aligned}
 \|e_{i}(k)\|^{2}\ge &  \frac{c_i}{d_{in}^{i}(k)}\left(\frac{1}{2}- \beta  d_{in}^{i}(k) \left(\frac{l_{i}^{2}}{\alpha^{2}}+T_{s}\frac{l_{i}}{\alpha}\right) \right)^{2}\cdot\\
 & \sum_{j=1}^{N} a_{ij}(k)\|\hat{\lambda}_{j}(k)-\hat{\lambda}_{i}(k)\| ^{2} \label{eq:event-driven condition 2}
\end{aligned}
\end{equation}
where $e_{i}(k)=\bar{\lambda}_{i}(k)-\hat{\lambda}_{i}(k)$ and $ c_i\in (0,1)$, then the algorithm
\eqref{eq:algorithm-1} with any initial condition with $\sum_{i=1}^{N}\gamma_{i}(0)=\mathbf{0}$
will converge to the optimal solution of \eqref{eq:distributed convex optimization}. \end{thm}
\textbf{Proof.} The proof is stated in  Appendix.\endprf }

\vspace{2mm}
\textcolor{black}{
Under the event triggering condition \eqref{eq:event-driven condition 2}, each node broadcasts its current state (after sampling) $\bar{\lambda}_i(k)$ to its out-neighbors when a local “error” signal exceeds a threshold depending on its own cost function and the last received state of $\hat{\lambda}_j(k)$ from its in-neighbors. Such an triggering condition requires   each node being  aware of the existence of its in-neighbors.  Whenever an edge between two nodes is established, the sender sends its last triggered state to the receiver, which is not considered as a “triggering”. Whenever an edge is canceled or established, the receiver updates its input $\hat{u}_i(k)$ by removing or adding the corresponding item of $\hat{\lambda}$.
\begin{rem}\label{re:event}
Given fixed $\alpha,~\beta$ and $T_s$, condition \eqref{eq:event-driven condition 2} is a simple and distributed one to be verified by each node over a balanced graph with very weak connectivity (Assumption~\ref{assum:3}).
It is worth mentioning that this sampled-based event-triggered communication scheme is exempt from Zeno behavior as the minimum inter-event time can be guaranteed by the sampling period $T_s$.
\end{rem}}

 \section{Simulation}\label{sec:4}
 In this section, a numerical example is provided to illustrate the previous results.

 Consider the resource allocation problem \eqref{eq:resource allocation problem} with $N=10, m=2$, and
 \[
 \begin{array}{ll}
      f_1(x_1)=x_{11}^2+\frac{1}{2}x_{11}x_{12}+\frac{1}{2}x_{12}^2+1; \quad f_2(\cdot)=f_1(\cdot);\\
      f_3(x_3)=\frac{1}{4}(x_{31}+2)^2+x_{32}^2;  \quad f_4(\cdot)=f_3(\cdot); \\
      f_5(x_5)=\frac{1}{2} x_{51}^{2}-\frac{1}{2}x_{51}x_{52}+x_{52}^{2};  \quad f_6(\cdot)=f_5(\cdot);\\
      f_7(x_7)=\ln (e^{2x_{71}}+1)+x_{72}^{2};  \quad f_8(\cdot)=f_7(\cdot);\\
      f_9(x_9)=\ln (e^{2x_{91}}+e^{-0.2x_{91}})+\ln (e^{x_{92}}+1); ~ f_{10}(\cdot)=f_9(\cdot).  
 \end{array}
 \]
 and $d_1=d_2=d_3=d_4=d_5=[1 \; 1]^{T}$, $d_6=d_7=d_8=d_9=d_{10}=[2 \; 2]^{T}$. Suppose the communication graph $\mathcal{G}(t)$ is time varying, which alternates every 1s between $\mathcal{G}_1$ and $\mathcal{G}_2$ shown in Fig. 2. 
 It can be observed that the switching graph $\mathcal{G}(t)$ is weight-balanced and infinitely jointly strongly connected, and  Assumption  \ref{assum:1} holds with $l_1=l_2=l_5=l_6=2.21, l_3=l_4=1_7=l_8=2,l_9=l_{10}=1.21$.
 
 \begin{figure}[tbh]
  \centering
  \includegraphics[width=1\linewidth]{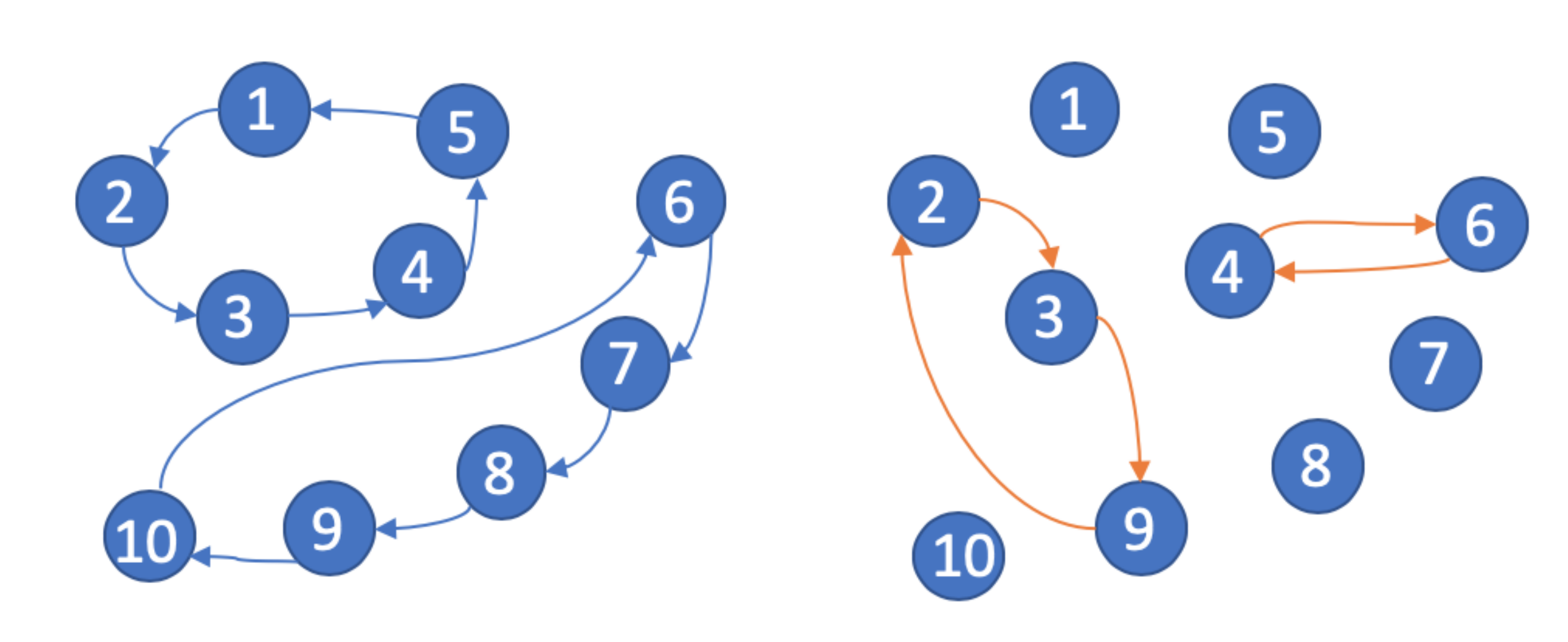}
  \center
  \caption{The switching communication graph $\mathcal{G}(t)$.}
  \label{fig:G}
\end{figure}

We solve the centralized convex problem \eqref{eq:resource allocation problem} using Yalmip, and obtain the optimal solution $x_{i}^{*},i=1,\ldots,10$. According to Lemma \ref{lem:0}, $\lambda_1^{*}=\ldots=\lambda_{10}^{*}=\nabla f_i(x_i^{*})=[1.87 \; 0.992]^{T}$.   The goal is to design a continuous-time distributed algorithm to equivalently solve the optimization problem \eqref{eq:resource allocation problem} under discrete-time communication.

To start with, we recast the above problem into  \eqref{eq:distributed convex optimization} based on  Section \ref{sub:The-Lagrange-dual}. It can be obtained that $\Delta J_i(\lambda_i)=h_i(\lambda_i)-d_i$ with
\[
\begin{array}{ll}
h_{1}(\lambda_{1})=\begin{pmatrix}\begin{array}{l}
\frac{4}{7}\lambda_{11}-\frac{2}{7}\lambda_{12}\\
\frac{8}{7}\lambda_{12}-\frac{2}{7}\lambda_{11}
\end{array}\end{pmatrix}; & h_{2}(\cdot)=h_{1}(\cdot);\\
h_{3}(\lambda_{3})=\begin{pmatrix}\begin{array}{l}
2\lambda_{31}-2\\
\frac{1}{2}\lambda_{32}
\end{array}\end{pmatrix}; & h_{4}(\cdot)=h_{3}(\cdot)\\
h_{5}(\lambda_{5})=\begin{pmatrix}\begin{array}{l}
\frac{8}{7}\lambda_{51}+\frac{2}{7}\lambda_{52}\\
\frac{2}{7}\lambda_{51}+\frac{4}{7}\lambda_{52}
\end{array}\end{pmatrix}; & h_{6}(\cdot)=h_{5}(\cdot);\\
h_{7}(\lambda_{7})=\begin{pmatrix}\begin{array}{l}
\frac{1}{2} \ln \frac{\lambda_{71}}{2-\lambda_{71}}\\
\frac{1}{2}\lambda_{72}
\end{array}\end{pmatrix} ;& h_{8}(\cdot)=h_{7}(\cdot);
\end{array}
\]
 \[
\begin{array}{ll}
h_{9}(\lambda_{9})=\begin{pmatrix}\begin{array}{l}
\frac{5}{11} \ln \frac{5\lambda_{91}+1}{10-5\lambda_{91}}\\
\ln\frac{\lambda_{91}}{1-\lambda_{91}}
\end{array}\end{pmatrix}; & h_{10}(\cdot)=h_{9}(\cdot).
\end{array}
\]
In the following simulations, we fix $\alpha=1$, and fix $\gamma_i(0)=\mathbf{0},\forall i\in \mathcal{I}$ to satisfy the initial condition  $\sum_{i=1}^{N}\gamma_i(0)=\mathbf{0}$. To examine the effectiveness of the distributed algorithms amounts to checking whether the trajectories of $\lambda_i(t),i\in\mathcal{I}$ converge to the value $\lambda^{*}=[1.87\; 0.992]^{T}$.

\begin{figure}[tbh]
    \centering
    \subfigure[]{\includegraphics[width=0.5\linewidth]{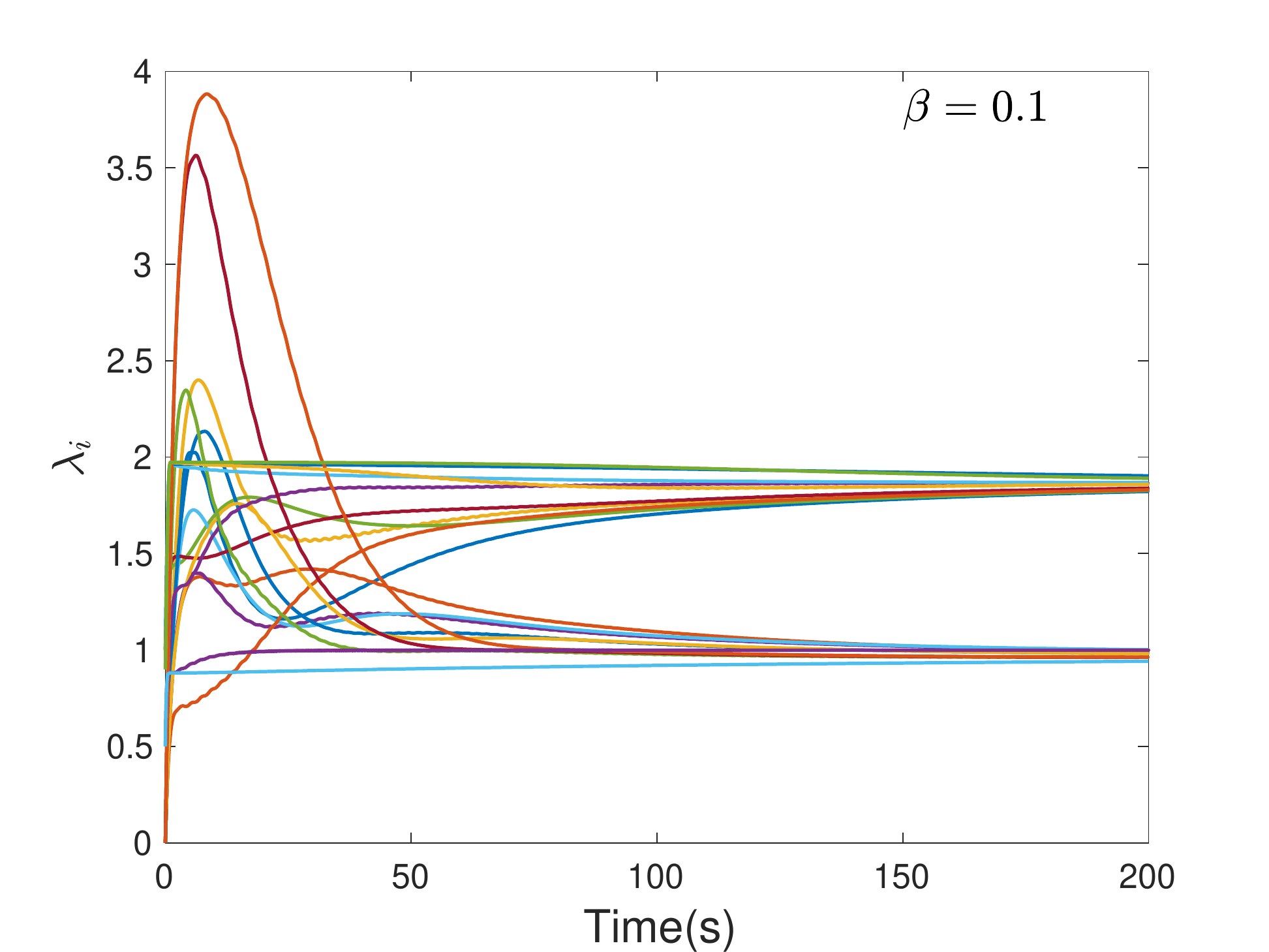}}\subfigure[]{\includegraphics[width=0.5\linewidth]{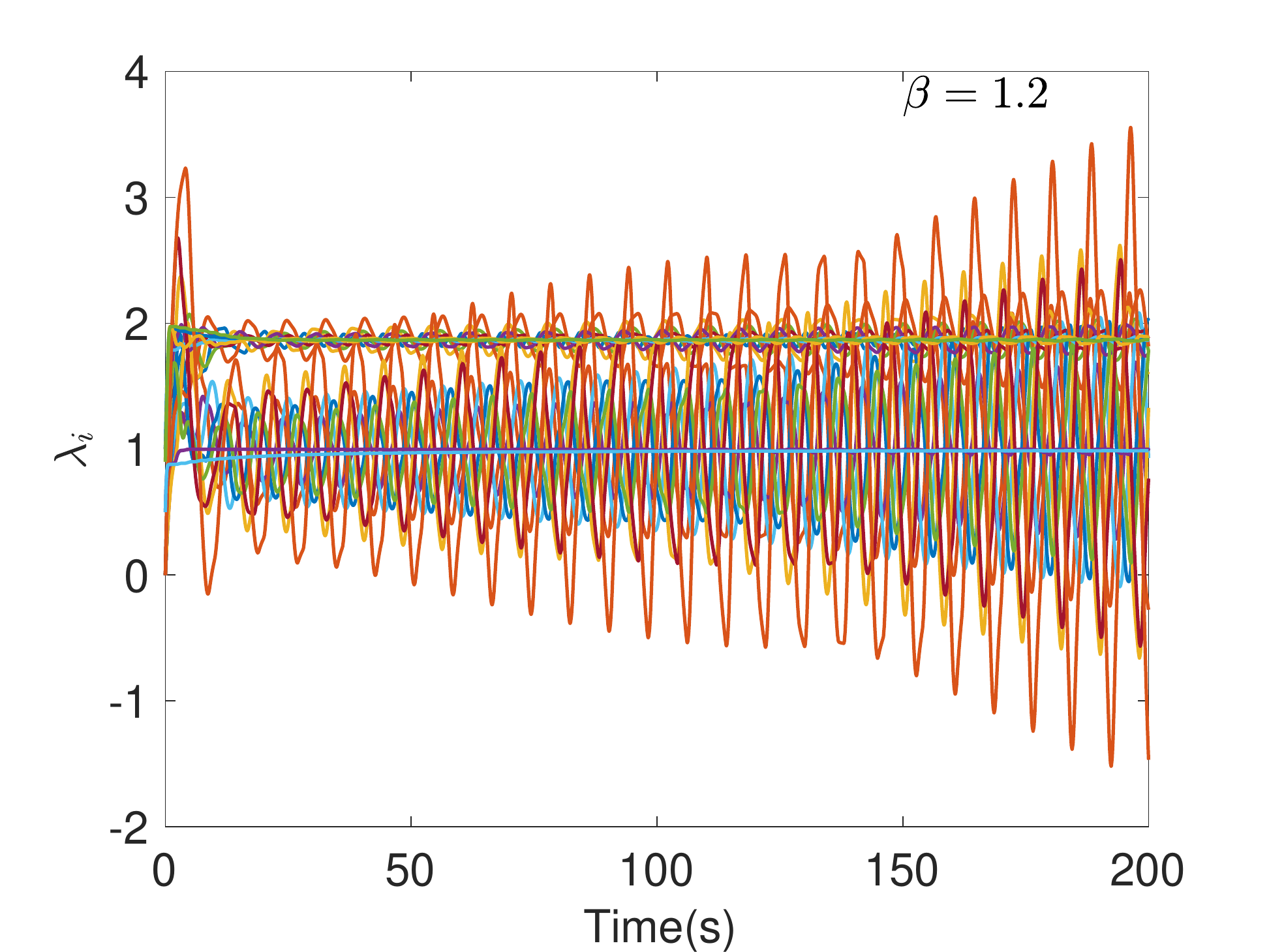}}
    \caption{Trajectories of $\lambda_i(t)$ under continuous communication.}
    \label{fig:result1}
\end{figure}

Let us first implement the distributed algorithm \eqref{eq:algorithm}  under continuous communication.   By the condition \eqref{eq: individual condition} in Corollary \ref{cor: conditions}, one has that the algorithm \eqref{eq:dynamic_compact form} will converge with $0 < \beta < 0.103$.  Under randomly generated initial value of $x_i(0)$, the trajectories of $\lambda_i(t), i\in\mathcal{I}$ are shown in Figure \ref{fig:result1} with different value of $\beta$. Although condition \eqref{eq: individual condition} is only sufficient, it is shown in Figure \ref{fig:result1} that the convergence is no longer ensured when $\beta$ takes some larger value. 

\begin{figure}[tbh]
    \centering
    \subfigure[]{\includegraphics[width=0.5\linewidth]{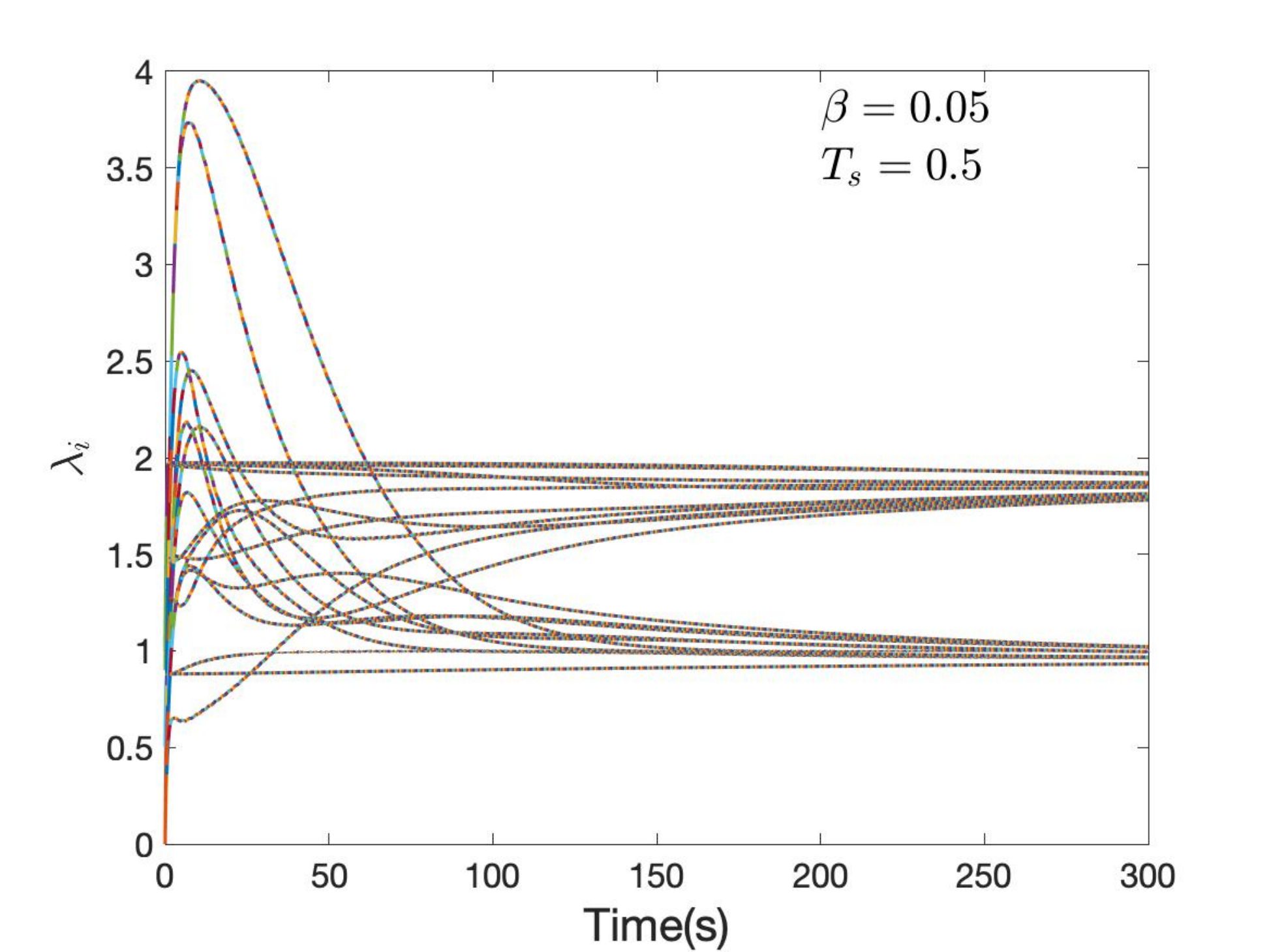}\label{fig:result3.1}}\subfigure[]{\includegraphics[width=0.5\linewidth]{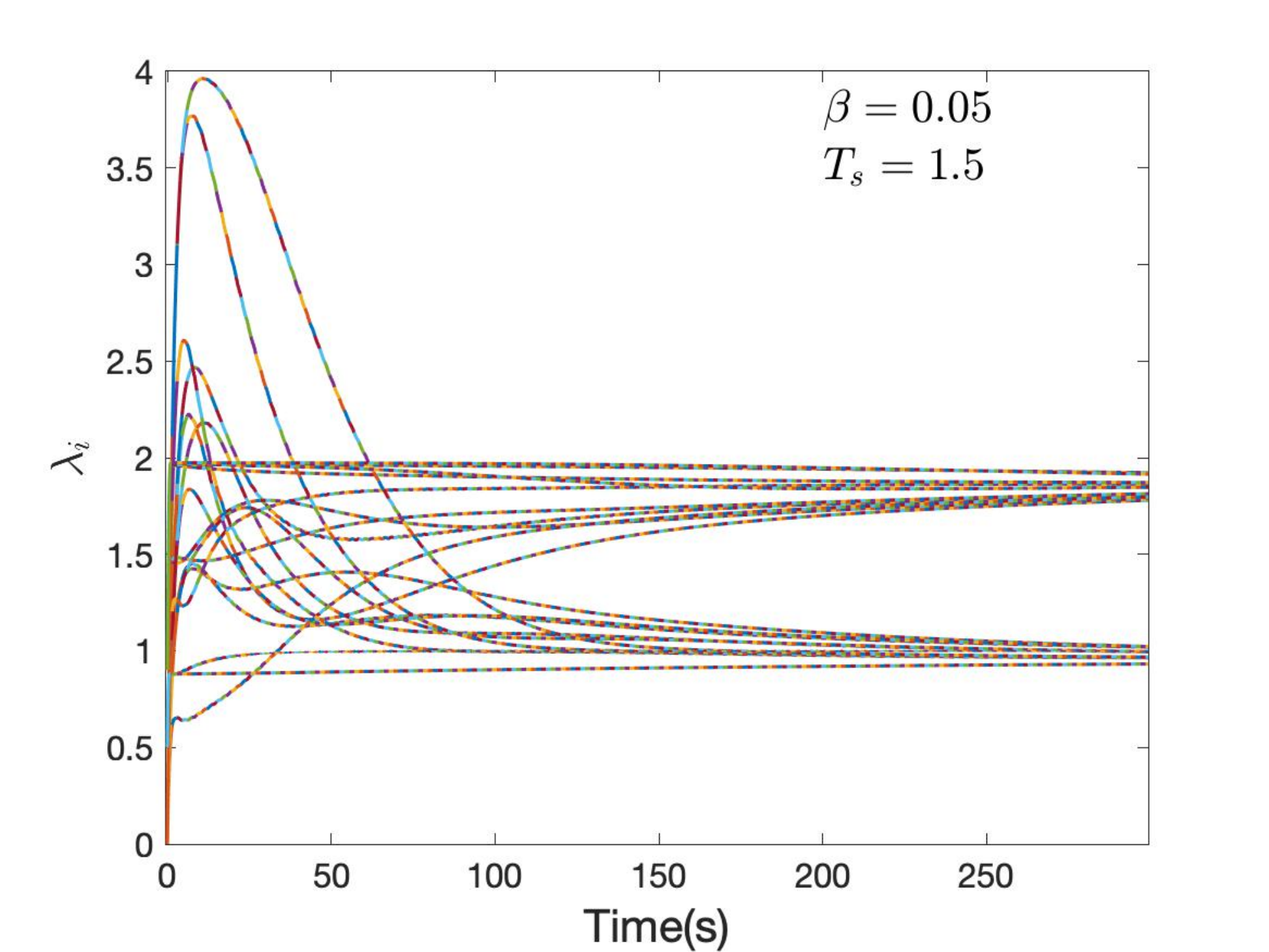}}
    \caption{Trajectories of $\lambda_i(t)$ under periodic communication.}
    \label{fig:result3}
\end{figure}

Next, we explore the distributed algorithm \eqref{eq:sampled algorithm} under periodic communication. By exploiting the condition \eqref{eq:beta_cond_sampling}, we have that the algorithm \eqref{eq:sampled algorithm} will converge with $0<\beta<\frac{1}{9.74 + 4.41 T_s}$. If we let $\beta=0.05$, then the condition yields that $T_s<2.3$. In this example, we let $T_s=0.5,1.5$ and it is obvious that Assumption \ref{assum 2} holds. The trajectories of $\lambda_i(t)$ are shown in Figure \ref{fig:result3}. \textcolor{black}{Note that the $T_s$ here is relatively large such that communication is greatly reduced.}

\begin{figure}[tbh]
    \centering
    \subfigure[]{\includegraphics[width=0.5\linewidth]{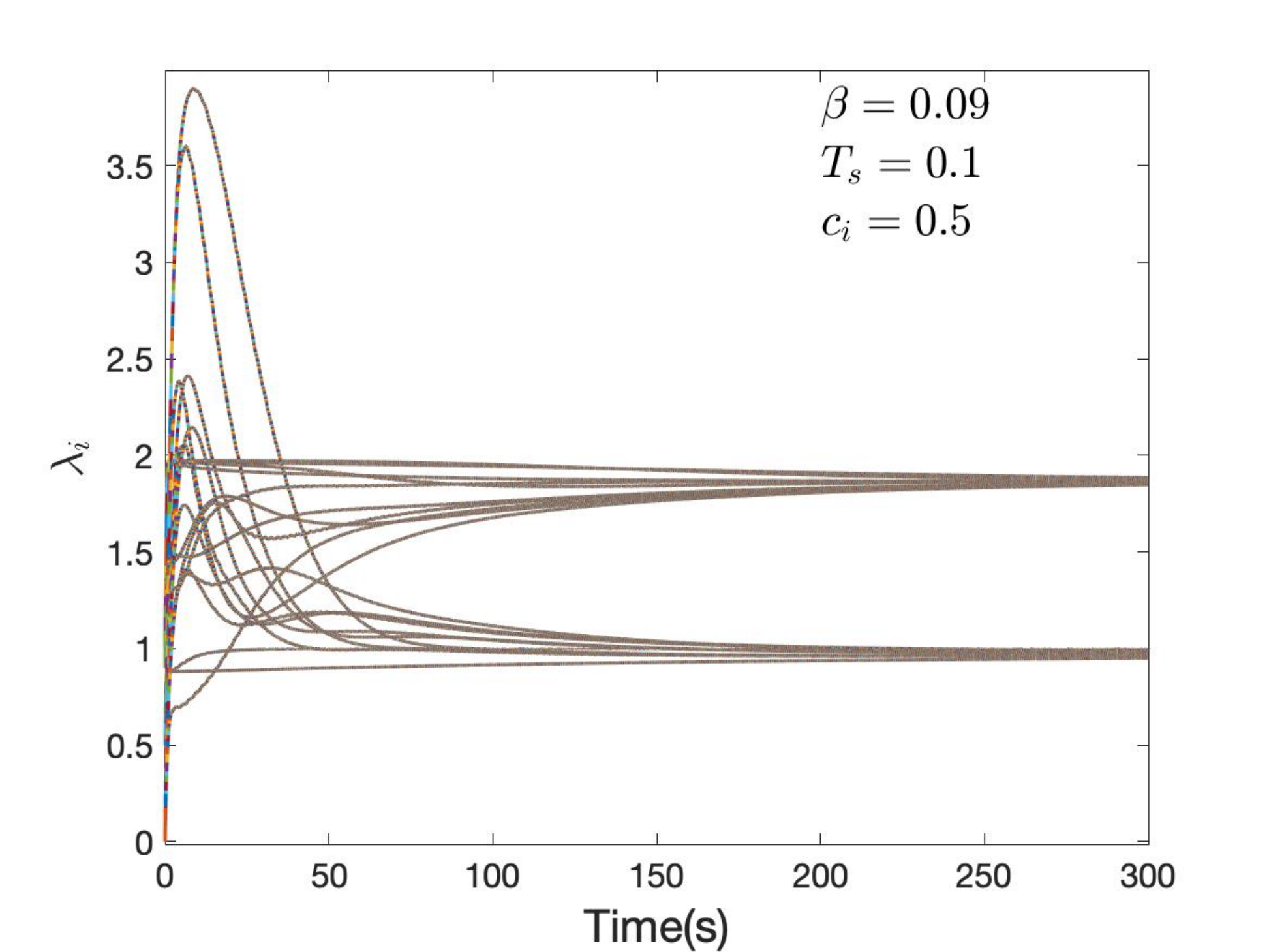} \label{fig:result4.1}}\subfigure[]{\includegraphics[width=0.5\linewidth]{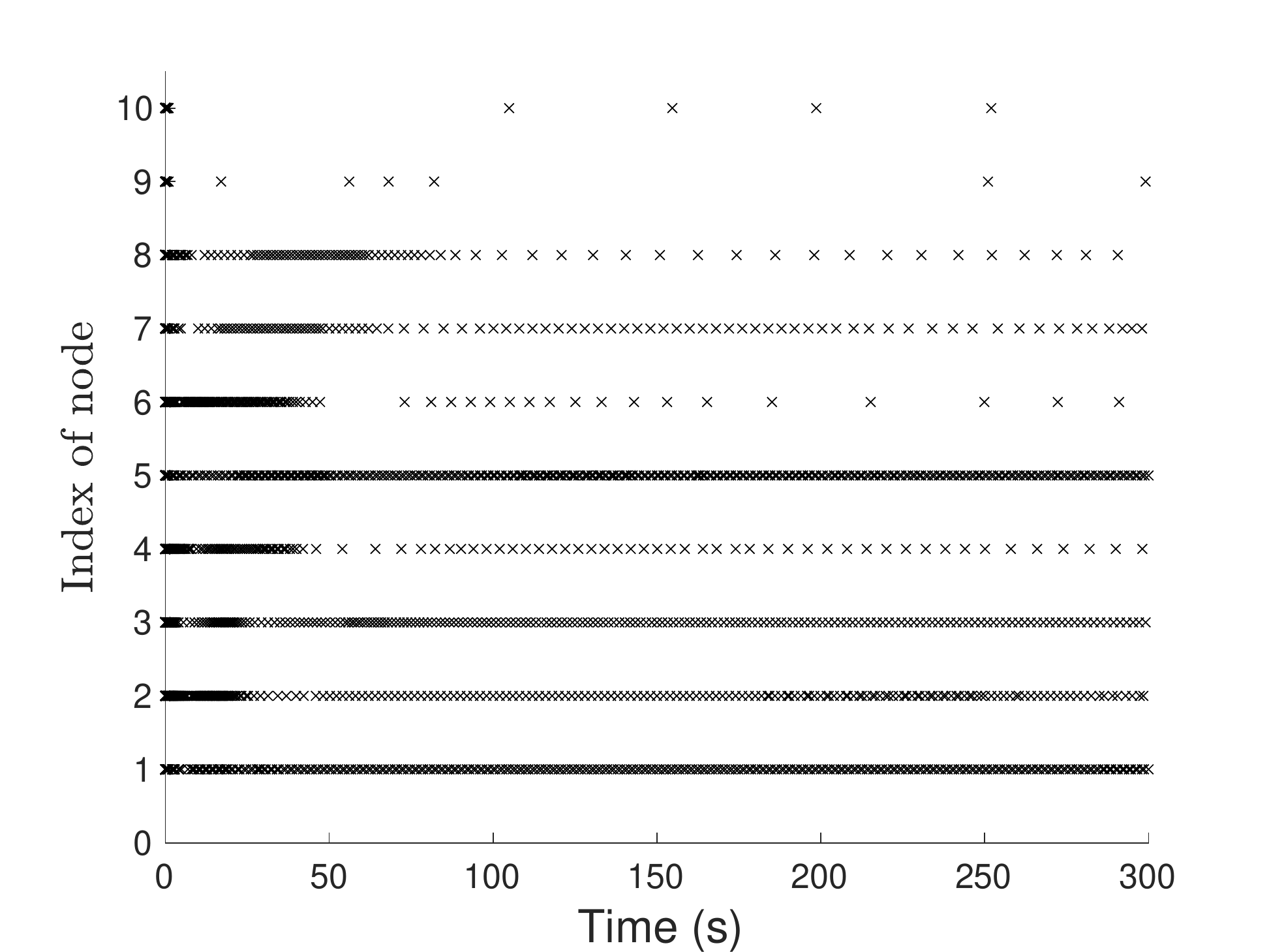} \label{fig:result4.2}}
    \caption{Trajectories and and triggering instants of $\lambda_i(t)$ under sample-based event-triggered communication.}
    \label{fig:result4}
\end{figure}

\textcolor{black}{
In the end, let us illustrate the distributed algorithm \eqref{eq:algorithm-1} under sample-based event-triggered communication. We select $\beta = 0.09$, $T_s = 0.1$ and $c_i = 0.5$ in \eqref{eq:event-driven condition 2}. The trajectories and the triggering instants of $\lambda_i(t)$ are shown in Figure~\ref{fig:result4}. In Figure~\ref{fig:result4.2}, the largest number of triggering times is $337$ for node $5$ while the smallest one is only $13$ for node $9$ and  node $10$, both of which are a lot smaller than the number of periodic sampling times $300 / T_s = 3000$. These shows that the sample-based event-triggered control effectively reduces communication costs. Moreover, a better convergence performance is observed in Figure~\ref{fig:result4.1} than the one in Figure~\ref{fig:result3.1} with less triggering times, due to the larger coupling gain $\beta$.}

\section{Conclusion}\label{sec:5}
\textcolor{black}{
We have introduced the passivity-based perspective for the continuous-time algorithm addressing  the distributed resource allocation problem over weight-balanced and infinitely jointly connected digraphs. By showing that the individual algorithmic dynamics is IFP, it is shown how to redesign the algorithm with intermittent communication protocol.} 

\textcolor{black}{The passivity-based analysis in this work is based on an existing algorithm that considers the distributed optimization without set constraints and with strong assumptions on cost functions. An interesting future direction is to investigate the passivity property for more advanced algorithms and apply it to time-varying graphs. Another promising direction is to explore the compatibility of the passivity-based approach with other network-induced imperfections such as time delay and packet drops. }

\appendix
\section*{Proof of Theorem \ref{thm:IFP}}
Since the Jacobian of $h_{i}(\lambda_{i})$ satisfies 
$
\frac{1}{l_{i}}I\le\frac{\partial h_{i}(\lambda_{i})}{\partial\lambda_{i}},
$ it follows from Mean Value Theorem that  
$h_{i}(\lambda_{i})-h_{i}(\lambda_{i}^{*})=B_{\lambda_{i}}\left(\lambda_{i}-\lambda_{i}^{*}\right)$ where  $B_{\lambda_i}$ is a symmetric  $\lambda_{i}$-dependent matrix defined as $B_{\lambda_{i}}=\int_{0}^{1}\frac{\partial h_i}{\partial \lambda_i}(\lambda_i+t(\lambda_i-\lambda_i^{*}))dt$
and $\frac{1}{l_{i}}I\le B(\lambda_{i})$.
Therefore, the system \eqref{eq: Psi} can be rewritten as 
\[
\left\{ \begin{array}{rcl}
\Delta\dot{\lambda}_{i} & = & -\alpha B_{\lambda_i}\Delta\lambda_{i}-\Delta\gamma_{i}\\
\Delta\dot{\gamma}_{i} & = & -u_{i}\\
u_{i} & = & \beta\sum_{j=1}^{N}a_{ij}(t)(\Delta\lambda_{j}-\Delta\lambda_{i}).
\end{array}\right.
\]
Consider the storage function 
\begin{equation}
\begin{array}{rcl}
V_{i} & = & \frac{\eta_i}{2}||\Delta\dot{\lambda}_{i}||^{2}-\Delta\lambda_{i}^{T}\Delta\gamma_{i}+\alpha(J_{i}(\lambda_{i}^{*})-J_{i}(\lambda_{i})\\
 &  & +\left(h_{i}(\lambda^{*})-d_{i}\right)^{T}\Delta\lambda_{i})
\end{array}\label{eq:storage function}
\end{equation}
where $\eta_i$ is chosen to satisfy $\eta_i>\frac{l_{i}}{\alpha}$. 

First, let us verify the positive definiteness of $V_{i}$.

It can be observed that $\frac{\eta_i}{2}||\Delta\dot{\lambda}_{i}||^{2}=\frac{\eta_i}{2}||\alpha B_{\lambda_i}\Delta\lambda_{i}+\Delta\gamma_{i}||^{2},$
and the strong convexity of $J_{i}(\lambda_i)$ provides that 
\[
J_{i}(\lambda_{i}^{*})-J_{i}(\lambda_{i})\ge-\left(h_{i}(\lambda_{i})-d_{i}\right)^{T}\Delta\lambda_{i}+\frac{1}{2l_{i}}||\Delta\lambda_{i}||^{2},
\]
which follows that the last term in the storage function $V_{i}$ \eqref{eq:storage function}
satisfies 
\[
\begin{array}{rl}
 & \alpha\left(J_{i}(\lambda_{i}^{*})-J_{i}(\lambda_{i})+\left(h_{i}(\lambda_i^{*})-d_{i}\right)^{T}\Delta\lambda_{i}\right)\\
\ge & \alpha\left(-\left(h_{i}(\lambda_{i})-h_{i}(\lambda_i^{*})\right)^{T}\Delta\lambda_{i}+\frac{1}{2l_{i}}||\Delta\lambda_{i}||^{2}\right)\\
= & \Delta\lambda_{i}^{T}\left(-\alpha B_{\lambda_i}+\frac{\alpha }{2l_{i}}I\right)\Delta\lambda_{i}.
\end{array}
\]
It can be derived that 
\begin{equation}
\begin{array}{l}
V_{i} \ge  \frac{\eta_i}{2}||\alpha B_{\lambda_i}\Delta\lambda_{i}+\Delta\gamma_{i}||^{2}-\Delta\lambda_{i}^{T}\Delta\gamma_{i}\\ \;\;\;\;\;\; +(\frac{\alpha} {2l_{i}}I-\alpha B_{\lambda_i})||\Delta\lambda_{i}||^{2}\\  =
\begin{pmatrix}\Delta\lambda_{i}\\
\Delta\gamma_{i}
\end{pmatrix}^{T}\underset{{\scriptstyle W}}{\underbrace{\begin{pmatrix}\frac{\alpha^{2}\eta_i}{2} B_{\lambda_i}^{2} -\alpha B_{\lambda_i}+\frac{\alpha}{2l_{i}}I & *\\
\frac{\alpha\eta_i}{2} B_{\lambda_i}-\frac{1}{2}I & \frac{\eta_i}{2}I
\end{pmatrix}}}\begin{pmatrix}\Delta\lambda_{i}\\
\Delta\gamma_{i}
\end{pmatrix}.
\end{array}\label{eq:PSD_storage function}
\end{equation}
Since $\frac{\eta_i}{2}I>0$, $\eta_i>\frac{l_{i}}{\alpha }$
and $\frac{\alpha^{2}\eta_i}{2}B_{\lambda_{i}}^{2}-\alpha B_{\lambda_{i}}+\frac{\alpha} {2l_{i}}I-\left(\frac{\alpha\eta_i}{2}B_{\lambda_{i}}-\frac{1}{2}I\right)\left(\frac{\eta_i}{2}I\right)^{-1}\left(\frac{\alpha\eta_i}{2}B_{\lambda_{i}}-\frac{1}{2}I\right)=-\frac{1}{2\eta_i}I+\frac{\alpha} {2l_{i}}I>0,$
it can be concluded based on Schur Complement Lemma that $W>0$. Therefore,
it can be claimed that $V_{i}\ge0$ and $V_{i}=0$ if and only if
$(\lambda_{i},\gamma_{i})=(\lambda_{i}^{*},\gamma_{i}^{*})$.

The next step is to show that with the defined storage function $V_{i}$,
the system $\text{\ensuremath{\Psi}}_{i}$ is IFP($\nu_{i}$) from
$u_{i}$ to $\Delta\lambda_{i}$. 

Let us observe that 
\[
\begin{array}{rcl}
\frac{\eta_i}{2}\cdot\frac{d||\Delta\dot{\lambda}_{i}||^{2}}{dt} & = & \eta_i\Delta\dot{\lambda}_{i}^{T}\left(-\alpha\frac{dh_{i}(\lambda_{i})}{dt}-\Delta\dot{\gamma}_{i}\right)\\
 & = & \eta_i\Delta\dot{\lambda}_{i}^{T}\left(-\alpha\frac{\partial h_{i}(\lambda_{i})}{\partial\lambda_{i}}\Delta\dot{\lambda}_{i}+u_{i}\right)\\
 & \le & -\frac{\eta_i\alpha }{l_{i}}||\Delta\dot{\lambda}_{i}||^{2}+\eta_i\Delta\dot{\lambda}_{i}^{T}u_{i},\\
 \frac{d(-\Delta\lambda_{i}^{T}\Delta\gamma_{i})}{dt}&=&-\Delta\dot{\lambda}_{i}^{T}\Delta\gamma_{i}+\Delta\lambda_{i}^{T}u_{i}.
\end{array}
\]
Recall that  $\nabla J_{i}(\lambda_{i})=h_{i}(\lambda_{i})-d_{i}$, and it follows 
\begin{eqnarray*}
 &  & \alpha\cdot\frac{d\left(J_{i}(\lambda_{i}^{*})-J_{i}(\lambda_{i})+\left(h_{i}(\lambda_{i}^{*})-d_{i}\right)^{T}\Delta\lambda_{i}\right)}{dt}\\
 & = & \alpha\left(-\nabla J_{i}(\lambda_{i})+\left(h_{i}(\lambda_{i}^{*})-d_{i}\right)\right)^{T}\Delta\dot{\lambda}_{i}\\
 & = & -\left(\alpha B_{\lambda_{i}}\Delta\lambda_{i}\right)^{T}\Delta\dot{\lambda}_{i}.
\end{eqnarray*}
By combining the above equations, one has that 
\begin{eqnarray}
\dot{V_{i}} & = & \frac{\eta_i}{2}\cdot\frac{d||\Delta\dot{\lambda}_{i}||^{2}}{dt}+\frac{d(-\Delta\lambda_{i}^{T}\Delta\gamma_{i})}{dt}+\nonumber \\
 &  & \alpha\cdot\frac{d\left(J_{i}(\lambda_{i}^{*})-J_{i}(\lambda_{i})+\left(h_{i}(\lambda_{i}^{*})-d_{i}\right)^{T}\Delta\lambda_{i}\right)}{dt}\nonumber \\
 & \le & -\frac{\eta_i\alpha }{l_{i}}||\Delta\dot{\lambda}_{i}||^{2}+\eta_i\Delta\dot{\lambda}_{i}^{T}u_{i}+\Delta\lambda_{i}^{T}u_{i}\nonumber \\
 &  & -\left(\alpha B(\lambda_{i})\Delta\lambda_{i}+\Delta\gamma_{i}\right)^{T}\Delta\dot{\lambda}_{i}\nonumber \\
 & = & \left(-\frac{\eta_i\alpha }{l_{i}}+1\right)||\Delta\dot{\lambda}_{i}||^{2}+\eta_i\Delta\dot{\lambda}_{i}^{T}u_{i}+\Delta\lambda_{i}^{T}u_{i}\label{eq:ifp proof}
\end{eqnarray}
with $-\frac{\eta_i\alpha }{l_{i}}+1<0$. Since 
\[
\left(-\frac{\eta_i\alpha }{l_{i}}+1\right)||\Delta\dot{\lambda}_{i}||^{2}+\eta_i\Delta\dot{\lambda}_{i}^{T}u_{i}\le\frac{\eta_i^{2}}{4\left(\frac{\eta_i\alpha}{l_{i}}-1\right)}u_{i}^{T}u_{i},
\]
 it follows that 
\[
\dot{V_{i}}\le\Delta\lambda_{i}^{T}u_{i}+\frac{\eta_i^{2}}{4\left(\frac{\eta_i\alpha }{l_{i}}-1\right)}u_{i}^{T}u_{i}.
\]
 Finally, let us prove $\nu_{i}\ge-\frac{l_{i}^{2}}{\alpha^{2}}.$
To this end, consider the following optimization problem
\[
\underset{\eta_i>\frac{l_{i}}{\alpha }}{\text{min}}\ \frac{\eta_i^{2}}{4\left(\frac{\eta_i\alpha }{l_{i}}-1\right)},
\]
and it can be verified that the optimal solution is given by $\eta_i^{*}=\frac{2l_{i}}{\alpha}$
and the corresponding minimum value of the above objective function
is $\frac{l_{i}^{2}}{\alpha^{2}}$. 

Thus, it can be summarized that $\dot{V_{i}}\le\Delta\lambda_{i}^{T}u_{i}+\frac{l_{i}^{2}}{\alpha^{2}}u_{i}^{T}u_{i}$,
which completes the proof.

\vspace{2mm}
\section*{Proof of Theorem \ref{thm:convergence}}
Recall the storage function defined in \eqref{eq:storage function}
for individual system, and consider the Lyapunov function $V=\sum_{i=1}^{N}V_{i}$
for the overall distributed algorithm. Denote $u=\text{col}(u_{1},\ldots,u_{N})$,
$\Delta\lambda=\text{col}(\Delta\lambda_{1},\ldots,\Delta\lambda_{N})$,
and it follows from \eqref{eq: Psi} that $u=-\beta\left(L(t)\otimes I_{m}\right)\Delta\lambda$.
Based on the result in Theorem \ref{thm:IFP}, one has 
\[
\begin{array}{l}
\dot{V}\le\sum_{i=1}^{N}\Delta\lambda_{i}^{T}u_{i}+\frac{l_{i}^{2}}{\alpha^{2}}u_{i}^{T}u_{i}\\
=-\beta\Delta\lambda^{T}\left(L(t)\otimes I_{m}\right)\Delta\lambda+\beta^{2}\Delta \lambda^{T}\left(L(t)^{T}\otimes I_m\right)\times\\\;\;\;\left(\text{diag}\left(\frac{l_{i}^{2}}{\alpha^{2}}\right)\otimes I_m\right)\left(L(t)\otimes I_{m}\right)\Delta\lambda\\
=\Delta\lambda^{T}\left(M\otimes I_{m}\right)\Delta\lambda
\end{array}\label{eq:consensus proof}
\]
with
\[M=-\frac{\beta}{2}\left(L(t)+L(t)^{T}\right)+\beta^{2}\left(L(t)^{T}\text{diag}\left(\frac{l_{i}^{2}}{\alpha^{2}}\right)L(t)\right).\]

Since a weight-balanced digraph $\mathcal{G}$ is strongly connected if and only if it is weakly connected (Lemma 1 in \cite{chopra2006passivity}), any weight-balanced digraph amounts to the union of a set of strongly connected balanced graphs. For a strongly connected balanced graph, it is apparent that its Laplacian $L$ has the same null space with $L^{T}$, which is $\text{span}\{1_N\}$. Then, for a weight-balanced digraph, its Laplacian $L$ and $L^{T}$ have the same null space. Therefore,   $\text{Null}(L(t)+L(t)^{T})$ is the same with $\text{Null}(L(t)^{T}\text{diag}\left(l_{i}^{2}\right)L(t))$ at any time $t$. Besides,  since $\mathcal{G}(t)$ is weight-balanced for all $t$, it can be easily verified that $L(t)+L(t)^{T}\ge0$ and $L(t)^{T}\text{diag}\left(l_{i}^{2}\right)L(t)\ge0$. Since the above two matrices are both positive semi-definite and have the same null space, it can be implied from the min-max theorem that if the condition in \eqref{eq:beta_condition} holds, then 
\begin{equation}
 \alpha^{2}\left(L(t)+L(t)^{T}\right)\ge 2\beta L(t)^{T}\text{diag}\left( l_{i}^{2}\right)L(t).   \label{eq:M psd}
\end{equation}
Thus, it can be concluded that $M\le0$, which leads to $\dot{V}\le 0$. Note that at any time t, $M$ has the same null space with $L(t)$'s, so $\dot{V}(t)=0$ only if the nodes belonging to  the same strongly connected subgraph reach output consensus. According to LaSalle's invariance principle, the trajectory $\Delta \lambda$ tends to the largest invariant set of $\{\Delta\lambda|\dot{V}(t)=0\}$. Moreover, since the graph $\mathcal{G}(t)$ is \textcolor{black}{infinitely} jointly strongly connected, one has that  $\Delta \lambda$ will converge to  the  set $\{\Delta \lambda|\Delta\lambda_{1}=\ldots=\Delta\lambda_{N}\}$.

According to \eqref{eq:PSD_storage function},  $V \geq 0$ and $V$ is radially unbounded, i.e., $V\rightarrow\infty$ as $||(\Delta \lambda^{T}, \Delta\gamma^{T})^{T}||\rightarrow\infty$. Since $\dot{V} \leq 0$, then $V$ is non-increasing, and the state is bounded, i.e., $\lambda$, $\gamma$ are bounded. Let us recall that $\Lambda_i \triangleq  \text{range}(\nabla f_i(x_i))$ with $x_i\in\mathbb{R}^{m}$, and $h_i(\nabla f_i(x_i))=x_i$. Let $\bar{\Lambda}_i$ be the boundary of the set $\Lambda_i$. Since $x_i\in\mathbb{R}^{m}$ is unbounded in our Problem \eqref{eq:resource allocation problem} and $f_i$ is strictly convex, then  $||h_i(\lambda_i)|| \rightarrow \infty$ when  $\lambda_i \rightarrow \bar{\Lambda}_i$. From the first line of \eqref{eq:algorithm}, this yields that $||\dot{\lambda}_i|| \rightarrow \infty$ when  $\lambda_i \rightarrow \bar{\Lambda}_i$  since $\gamma_i$ is bounded. Consequently, based on \eqref{eq:storage function}, $V \rightarrow \infty$, which contradicts the fact that $V$ is non-increasing. Therefore, for all $i\in\mathcal{I}$, the set $\Lambda_i$ is a positively invariant set of $\lambda_i$.

Next, let us show that $\dot{V}=0\Rightarrow\Delta\dot{\lambda}_{1}=\ldots=\Delta\dot{\lambda}_{N}=\textbf{0}$.
Since the inequality in \eqref{eq:beta_condition} is strict, it follows
that there exists a small enough scalar $\epsilon>0$ such that 
\begin{equation}
0<\beta<\frac{\alpha^{2}\sigma_{min}^{+}(L(t)+L(t)^{T})}{2\sigma_{N}\left(L(t)^{T}\text{diag}\left(l_{i}^{2}+\epsilon\right)L(t)\right)}.\label{eq:beta conditio_2}
\end{equation}
By substituting $\eta_i$ with $\eta_i^{*}=\frac{2l_{i}}{\alpha }$
in \eqref{eq:ifp proof}, we have
 \[
\dot{V_{i}}  \le  -||\Delta\dot{\lambda}_{i}||^{2}+\frac{2l_{i}}{\alpha}\Delta\dot{\lambda}_{i}^{T}u_{i}+\Delta\lambda_{i}^{T}u_{i}.
\] 
By completing the square, we further have
$
-||\Delta\dot{\lambda}_{i}||^{2}+\frac{2l_{i}}{\alpha}\Delta\dot{\lambda}_{i}^{T}u_{i}\le -\frac{\epsilon}{\left(l_{i}^{2}/\alpha ^{2}+\epsilon\right)}||\Delta\dot{\lambda}_{i}||^{2}+\left(\frac{l_{i}^{2}}{\alpha ^{2}}+\epsilon\right)u_{i}^{T}u_{i}.
$
Hence, 
\begin{equation}
\dot{V_{i}}  \le  -\frac{\epsilon}{\left(\frac{l_{i}^{2}}{\alpha ^{2}}+\epsilon\right)}||\Delta\dot{\lambda}_{i}||^{2}+\left(\frac{l_{i}^{2}}{\alpha ^{2}}+\epsilon\right)u_{i}^{T}u_{i}+\Delta\lambda_{i}^{T}u_{i}.\label{eq:small perturbation}
\end{equation}
Hence, by similar argument before, it follows
that $\dot{V}\le\Delta\lambda^{T}\left(\hat{M}\otimes I_{m}\right)\Delta\lambda-\sum_{i=1}^{N} \frac{\epsilon}{\left(l_{i}^{2}/\alpha ^{2}+\epsilon\right)}||\Delta\dot{\lambda}_{i}||^{2}$
where $\hat{M}=-\frac{\beta}{2}\left(L(t)+L(t)^{T}\right)+\beta^{2}L(t)^{T}\text{diag}\left(\frac{l_{i}^{2}}{\alpha^{2}}+\epsilon\right)L(t)$
and $\hat{M}\le0$. As a consequence, it can be concluded that
$\dot{V}\le0$ and $\dot{V}=0$ only if $\Delta\dot{\lambda}_{1}=\ldots=\Delta\dot{\lambda}_{N}=\textbf{0}$.

Because of the LaSalle's invariance principle, we have that $\Delta\dot{\lambda}\rightarrow\textbf{0}$
and $\Delta\lambda\rightarrow1_{N}\otimes s$ for some $s\in\mathcal{\mathbb{R}}^{m}$
as $t\rightarrow \infty$. Furthermore, by \eqref{eq: Psi}, one has $\Delta\dot{\gamma}\rightarrow\textbf{0}$
as $t\rightarrow \infty$. Thus, the states $\lambda,\gamma$
under the algorithm \eqref{eq:algorithm} will converge to an equilibrium
point. With the initial condition $\sum_{i=1}^{N}\gamma_{i}(0)=\textbf{0}$,
it follows from Lemma \ref{lem:1} that the algorithm \eqref{eq:algorithm}
will converge to the optimal solution of the problem \eqref{eq:distributed convex optimization}.

\vspace{2mm}
\section*{Proof of Corollary \ref{cor: conditions}}
 Define a vector variable $x=\text{col}(x_1,\ldots,x_N)^{T} \in \mathbb{R}^{mN}$ and it can be observed that $x^{T}(\mathbf{L}(t)+\mathbf{L}(t)^T)x(t)=2\sum_{i=1}^{N}x_i \sum_{j=1}^{N}a_{ij}(t)(x_i-x_j)=\sum_{i=1}^{N}\sum_{j=1}^{N}a_{ij}(t)(x_i-x_j)^2$ where the second equality follows from the balance of the graph $\mathcal{G}(t)$. Suppose the condition \eqref{eq: individual condition} holds, i.e., $\alpha^{2}>2\beta2 l_{i}^{2}d_{in}^{i}(t)$ for all $i\in \mathcal{I}$. Then, one has  
\[\begin{array}{rl}
 & \alpha^{2}x^{T}(\mathbf{L}(t)+\mathbf{L}(t)^T)x(t)  =  \alpha ^{2}\displaystyle \sum_{i=1}^{N}\sum_{j=1}^{N}a_{ij}(t)(x_i-x_j)^2 
\\
 & \ge  2\beta\displaystyle\sum_{i=1}^{N}l_{i}^{2}d_{in}^{i}(t)\sum_{j=1}^{N}a_{ij}(t)(x_i-x_j)^2.
\end{array} \] 
Since $d_{in}^{i}(t)=\sum_{j=1}^{N}a_{ij}(t)$, it follows from Cauchy-Schwartz inequality that $d_{in}^{i}(t)\sum_{j=1}^{N}a_{ij}(t)(x_i-x_j)^2\ge \left(\sum_{j=1}^{N} a_{ij}(t)(x_i-x_j)\right)^2$. This yields that 
\[
 \begin{array}{rl}
 & \displaystyle\sum_{i=1}^{N}l_{i}^{2}d_{in}^{i}(t)\sum_{j=1}^{N}a_{ij}(t)(x_i-x_j)^2\\  \ge& \displaystyle\sum_{i=1}^{N}l_{i}^{2}\left(\sum_{j=1}^{N}a_{ij}(t)(x_i-x_j)\right)^2\\ =&x^{T} \mathbf{L}(t)^{T}\text{diag}(l_i^2)\mathbf{L}(t) x(t).
\end{array}
\]
Hence, we have for all $x\in \mathbb{R}^{mN}$, $\alpha^{2}x^{T}(\mathbf{L}(t)+\mathbf{L}(t)^T)x(t)\ge 2\beta x^{T} \mathbf{L}(t)^{T}\text{diag}(l_i^2)\mathbf{L}(t) x(t)$, which is equivalent to \eqref{eq:M psd}. Following the same reasoning after \eqref{eq:M psd} will complete the proof.

\section*{Proof of Theorem \ref{thm:ifp index_sampled system}}
Let us consider a revised storage function $\bar{V}_{i}=\frac{1}{T_{s}}\left(V_{i}+\kappa||z_{i}||^{2}\right)$
with $V_{i}$ defined in \eqref{eq:storage function} and the coefficient
$\kappa>0$ will be designed later. The positive definiteness of $\bar{V}_{i}$
can be easily verified since $V_{i}$ is positive definite according
to the proof of Theorem \ref{thm:IFP} and $\kappa||z_{i}||^{2}\ge0$. 

Consider the difference of $\bar{V}_{i}$ between two consecutive sampling
instants, $kT_{s}$ and $(k+1)T_{s}$ for any $k\in\mathbb{N}$, we
have
\[
\begin{array}{l}
\int_{kT_{s}}^{(k+1)T_{s}}\dot{\bar{V}}_{i}dt=\bar{V}_{i}((k+1)T_{s})-\bar{V}_{i}(kT_{s})=\\
\frac{1}{T_{s}}\left(\int_{kT_{s}}^{(k+1)T_{s}}\dot{V}_{i}dt+\kappa||z_{i}((k+1)T_{s})||^{2}-\kappa||z_{i}(kT_{s})||^{2}\right).
\end{array}
\]
It is proved by Theorem \ref{thm:IFP} that $\dot{V}_{i}\le\Delta\lambda_{i}^{T}u_{i}+\frac{l_{i}^{2}}{\alpha^{2}}u_{i}^{T}u_{i}$.
By expressing $\Delta\lambda_{i}(t)$ as $\Delta\bar{\lambda}_{i}(k)+\left(\Delta\lambda_{i}(t)-\Delta\bar{\lambda}_{i}(k)\right)$,
one has 
\[
\begin{array}{cl}
 & \int_{kT_{s}}^{(k+1)T_{s}}\dot{V}_{i}dt\\
\le & \int_{kT_{s}}^{(k+1)T_{s}}\Delta\bar{\lambda}_{i}(k)^{T}u_{i}dt+\int_{kT_{s}}^{(k+1)T_{s}}\\
 & \left(\Delta\lambda_{i}(t)-\Delta\bar{\lambda}_{i}(k)\right)^{T}u_{i}dt+\frac{l_{i}^{2}}{\alpha^{2}}\int_{kT_{s}}^{(k+1)T_{s}}u_{i}^{T}u_{i}dt\\
\le & T_{s}\Delta\bar{\lambda}_{i}(k)^{T}\bar{u}_{i}(k)+T_{s}\frac{l_{i}^{2}}{\alpha^{2}}||\bar{u}_{i}(k)||^{2}\\ & +\int_{kT_{s}}^{(k+1)T_{s}} \left(\frac{1}{2\theta}||\Delta\lambda_{i}(t)-\Delta\bar{\lambda}_{i}(k)||^{2}+\frac{\theta}{2}||\bar{u}_{i}(k)||^{2}\right)dt
\end{array}
\]
where $\theta$ can be any positive scalar, and the second inequality
holds since $u_{i}(t)$ is set to be a piecewise signal due to the
zero order holder \eqref{eq:sampling}. Lemma \ref{lem:sampled error}
provides 
\[
\begin{array}{ll}
 & \int_{kT_{s}}^{(k+1)T_{s}}||\Delta\lambda_{i}(t)-\Delta\bar{\lambda}_{i}(k)||^{2}dt\\
\le & T_{s}^{3}\frac{l_{i}^{2}}{\alpha ^{2}}||\bar{u}_{i}||^{2}+T_{s}^{2}\frac{l_{i}}{\alpha}\left(||z_{i}(kT_{s})||^{2}-||z_{i}((k+1)T_{s})||^{2}\right)
\end{array}
\]
which follows that 
\[
\begin{array}{cl}
 & \int_{kT_{s}}^{(k+1)T_{s}}\dot{V}_{i}dt\\
\le & T_{s}\Delta\bar{\lambda}_{i}(k)^{T}\bar{u}_{i}(k)+\frac{T_{s}l_{i}^{2}}{\alpha^{2}}||\bar{u}_{i}(k)||^{2}+\left(\frac{T_{s}\theta}{2}+\frac{T_{s}^{3}l_{i}^{2} }{2\theta\alpha ^{2}}\right)\cdot\\
 & ||\bar{u}_{i}(k)||^{2}+\frac{T_{s}^{2}l_i}{2\theta \alpha}\left(||z_{i}(kT_{s})||^{2}-||z_{i}((k+1)T_{s})||^{2}\right).
\end{array}
\]
By selecting $\theta$ to minimize the value of $\left(\frac{T_{s}\theta}{2}+\frac{T_{s}^{3}l_{i}^{2}}{2\theta \alpha ^{2}}\right)$,
it can be easily obtained that 
\[
\theta^{*}=T_{s}\frac{l_{i}}{\alpha}\text{ and }\min\left(\frac{T_{s}\theta}{2}+\frac{T_{s}^{3}}{2\theta}\frac{l_{i}^{2}}{\alpha ^{2}}\right)=T_{s}^{2}\frac{l_{i}}{\alpha}
\]
Now, let us choose $\theta=T_{s}\frac{l_{i}}{\alpha}$ and $\kappa=\frac{T_{s}}{2}$.
It follows that 
\[
\begin{array}{rl}
 & \bar{V}_{i}((k+1)T_{s})-\bar{V}_{i}(kT_{s})\\
= & \frac{1}{T_{s}}\left(\int_{kT_{s}}^{(k+1)T_{s}}\dot{V}_{i}dt+\kappa||z_{i}((k+1)T_{s})||^{2}-\kappa||z_{i}(kT_{s})||^{2}\right)\\
\le & \Delta\bar{\lambda}_{i}(k)^{T}\bar{u}_{i}(k)+\left(\frac{l_{i}^{2}}{\alpha^{2}}+T_{s}\frac{l_{i}}{\alpha}\right)||\bar{u}_{i}(k)||^{2}.
\end{array}
\]

Thus, it can be observed that the sampled system $\bar{\Psi}_{i}$
is IFP$(\bar{\nu}_{i})$ from $\bar{u}_{i}$ to $\Delta\bar{\lambda}_{i}$
with IFP index $\bar{\nu}_{i}\ge -\left(\frac{l_{i}^{2}}{\alpha^{2}}+T_{s}\frac{l_{i}}{\alpha}\right)$.

\textcolor{black}{\section*{Proof of Theorem \ref{thm:event}}}
\textcolor{black}{
First, let us consider the equilibrium point of \eqref{eq:algorithm-1}
with initial condition satisfying $\sum_{i=1}^{N}\gamma_{i}(0)=\mathbf{0}$
whose compact form is represented as 
\begin{equation}
 \begin{array}{rcl}
\dot{\lambda^{*}} & = & -\alpha(h(\lambda^{*})-d)-\gamma^{*}=\mathbf{0}\\
\dot{\gamma^{*}} & = & \beta\mathbf{L}(k)\hat{\lambda}^{*}=\mathbf{0}.
\end{array}   \label{eq: equilibrium_event}
\end{equation}
}

\textcolor{black}{By similar reasoning in Lemma  \ref{lem:1}, we can obtain
that $\sum_{i=1}^{N}\gamma_{i}(t)=\mathbf{0}$ for any $t>0$ and $\nabla J(\lambda^{*})=0$.
Besides, $\dot{\gamma^{*}}=\beta\mathbf{L}(t)\hat{\lambda}^{*}=\mathbf{0}$ leads
to $\hat{\lambda}_{i}^{*}=\hat{\lambda}_{j}^{*},\forall i,j\in\mathcal{I}$.
Due to the triggering condition \eqref{eq:event-driven condition 2},
we have $\| \lambda_{i}^{*}-\hat{\lambda}_{i}^{*}\| =0$,
indicating $\lambda^{*}=\hat{\lambda}^{*}$ and $\lambda_{i}^{*}=\lambda_{j}^{*},\forall i,j\in\mathcal{I}$.
Under Assumption \ref{assum:1}, the equilibrium $(\lambda^{*},\gamma^{*})$ is
unique with $\lambda^{*}$ being the optimal solution of \eqref{eq:distributed convex optimization}. }

\textcolor{black}{ Next, the error dynamics in each individual subsystem is obtained by comparing \eqref{eq:algorithm-1} and \eqref{eq: equilibrium_event} as 
\[
\hat{\Psi}_{i}:\left\{ \begin{array}{rcl}
\Delta \dot{\lambda_{i}} & = & -\alpha\left(h_{i}(\lambda_{i})-h_{i}(\lambda^{*})\right)-\Delta\gamma_{i}\\
\Delta \dot{\gamma_{i}} & = & -u_{i}\\
\hat{u}_{i} (k)& = & \beta\sum_{j=1}^{N}a_{ij}(k)(\Delta\hat{\lambda}_{j}(k)-\Delta\hat{\lambda}_{i}(k))
\end{array}\right.
\]
with $\Delta\hat{\lambda}_{i}=\hat{\lambda}_{i}-\lambda_{i}^{*}$.
Since the dynamic from input $u_{i}$ to output $\Delta\bar{\lambda}_{i}$
is the same with that in \eqref{eq:sampling Psi} and $u_{i}(t) = \hat{u}_{i}(k),\forall t\in[kT_{s},(k+1)T_{s})$, it follows from Theorem
\ref{thm:ifp index_sampled system} that 
\[
\begin{array}{rl}
 & \bar{V}_{i}((k+1)T_{s})-\bar{V}_{i}(kT_{s})\\
\le & \Delta\bar{\lambda}_{i}(k)^{T}\hat{u}_{i}(k)+\left(\frac{l_{i}^{2}}{\alpha^{2}}+T_{s}\frac{l_{i}}{\alpha}\right)||\hat{u}_{i}(k)||^{2}, \forall i\in\mathcal{I}.
\end{array}
\]
with $\bar{V}_{i}$ defined in the proof of Theorem \ref{thm:ifp index_sampled system}. Consider the Lyapunov
function $\bar{V}=\sum_{i=1}^{N}\bar{V}_{i}$, and it yields that 
\begin{align*}
& \bar{V}(k+1)-\bar{V}(k)\\
\le & { \displaystyle\sum_{i=1}^{N}}\Delta\bar{\lambda}_{i}(k)^{T}\hat{u}_{i}(k)+\left(\frac{l_{i}^{2}}{\alpha^{2}}+T_{s}\frac{l_{i}}{\alpha}\right)||\hat{u}_{i}(k)||^{2}\\
= &  { \displaystyle\sum_{i=1}^{N}}\beta\Delta\bar{\lambda}_{i}^{T} (k)\displaystyle\sum_{j=1}^{N}a_{ij}(k)\left(\Delta\hat{\lambda}_{j}(k)-\Delta\hat{\lambda}_{i}(k)\right) \\
& +{ \displaystyle\sum_{i=1}^{N}} \beta^{2}\left(\frac{l_{i}^{2}}{\alpha^{2}}+T_{s}\frac{l_{i}}{\alpha}\right) \left\Vert \displaystyle\sum_{j=1}^{N}a_{ij}(k)\left(\Delta\hat{\lambda}_{j}(k)-\Delta\hat{\lambda}_{i}(k)\right)\right\Vert^{2}\\
= & { \displaystyle\sum_{i=1}^{N}}\beta\left(\Delta\hat{\lambda}_{i}(k)+e_{i}(k)\right)^{T}{ \displaystyle\sum_{j=1}^{N}}a_{ij}(k)\left (\Delta\hat{\lambda}_{j}(k)-\Delta\hat{\lambda}_{i}(k)\right)\\
& + { \displaystyle\sum_{i=1}^{N}}\beta^{2}\left(\frac{l_{i}^{2}}{\alpha^{2}}+T_{s}\frac{l_{i}}{\alpha}\right) \left\Vert { \displaystyle\sum_{j=1}^{N}}a_{ij}(k)\left(\Delta\hat{\lambda}_{j}(k)-\Delta\hat{\lambda}_{i}(k)\right)\right\Vert ^{2}\\
= & \beta {\displaystyle \sum_{i=1}^{N}}  {\displaystyle \sum_{j=1}^{N}}e_{i}(k)^{T}a_{ij}(k)\left(\Delta\hat{\lambda}_{j}(k)-\Delta\hat{\lambda}_{i}(k)\right)+\beta {\displaystyle \sum_{i=1}^{N}} {\displaystyle\sum_{j=1}^{N}} \\
& a_{ij}(k)\Delta\hat{\lambda}_{i}(k)^{T}\Delta\hat{\lambda}_{j}(k)
 - \beta {\displaystyle \sum_{i=1}^{N}} {\displaystyle \sum_{j=1}^{N}}a_{ij}(k)\Delta\hat{\lambda}_{i}(k)^{T}\Delta\hat{\lambda}_{i}(k) \\
& + {\displaystyle \sum_{i=1}^{N}} \beta^{2} \left(\frac{l_{i}^{2}}{\alpha^{2}}+
 T_{s}\frac{l_{i}}{\alpha}\right) \left\Vert {\displaystyle \sum_{j=1}^{N}}a_{ij}(k)\left(\Delta\hat{\lambda}_{j}(k)-\Delta\hat{\lambda}_{i}(k)\right)\right\Vert ^{2} 
\end{align*}
where the second equality holds since $e_{i}(k)=\bar{\lambda}_{i}(k)-\hat{\lambda}_{i}(k)=\Delta\bar{\lambda}_{i}(k)-\Delta\hat{\lambda}_{i}(k)$.
It can be derived from $\mathcal{G}(t)$ being balanced that
\[
\begin{array}{l}
  { \sum_{i=1}^{N}\sum_{j=1}^{N}}a_{ij}(k)\left(\Delta\hat{\lambda}_{i}(k)^{T}\Delta\hat{\lambda}_{j}(k)-\Delta\hat{\lambda}_{i}(k)^{T}\Delta\hat{\lambda}_{i}(k)\right)\\
=  -\frac{1}{2}{\sum_{i=1}^{N}\sum_{j=1}^{N}}a_{ij}(k)\|\Delta\hat{\lambda}_{j}(k)-\Delta\hat{\lambda}_{i}(k)\|^{2}.
\end{array}
\]
Let us observe that for all $\tau_i > 0$
\[
    \begin{array}{rl}
    &  e_{i}(k)^{T} a_{ij}(k)\left(\Delta\hat{\lambda}_{j}(k)-\Delta\hat{\lambda}_{i}(k)\right) \\
  \le   &   a_{ij}(k)\left(  \frac{1}{2\tau_i }\|e_i(k) \|^2+\frac{\tau_i }{2} \|\hat{\lambda}_{j}(k)-\Delta\hat{\lambda}_{i}(k)\|^2\right)   \end{array}
\]
and that 
\[
\begin{array}{rl}
     & \left\| \sum_{j=1}^{N}a_{ij}(k)\left(\Delta\hat{\lambda}_{j}(k)-\Delta\hat{\lambda}_{i}(k)\right)\right\| ^{2} \\
    \le  & d_{in}^{i}(k) \sum_{j=1}^{N}a_{ij}(k)\|\Delta\hat{\lambda}_{j}(k)-\Delta\hat{\lambda}_{i}(k)\|^{2}
\end{array}
\]
which can be proved by Cauchy-Schwartz inequality.
 With the above equations, we can now have for any $\tau_i > 0$ 
 \[
\begin{array}{rl}
 & \bar{V}(k+1)-\bar{V}(k)\\
\le & \displaystyle -\frac{\beta}{2}  \sum_{i=1}^{N}\sum_{j=1}^{N} a_{ij}(k) \left( \left(1-\tau_i -2\beta d_{in}^{i}(k) \left(\frac{l_{i}^{2}}{\alpha^{2}}+T_{s}\frac{l_{i}}{\alpha}\right) \right)\cdot \right.\\
&  \displaystyle \left.   \|\Delta\hat{\lambda}_{j}(k)-\Delta\hat{\lambda}_{i}(k)\|^{2}  -\frac{\|e_i(k)\|^2}{\tau_i} \right)\\
= & \displaystyle-\frac{\beta}{2}   \sum_{i=1}^{N}\left(    \left(1- \tau_i -2\beta d_{in}^{i}(k)  \left(\frac{l_{i}^{2}}{\alpha^{2}}+T_{s}\frac{l_{i}}{\alpha}\right) \right)\right. \sum_{j=1}^{N}\\
&\left. a_{ij}(k) \displaystyle \|\Delta\hat{\lambda}_{j}(k)-\Delta\hat{\lambda}_{i}(k)\|^{2}  - d_{in}^{i}(k)\frac{\|e_i(k)\|^2}{\tau_i }  \right).
\end{array}
\]
By letting $\tau_i =\frac{1}{2}-\beta d_{in}^{i}(k) \left(\frac{l_{i}^{2}}{\alpha^{2}}+T_{s}\frac{l_{i}}{\alpha}\right) $, it can be verified by \eqref{eq:beta_cond_sampling} that $\tau_i > 0$, and the above inequality becomes
 \[
\begin{array}{rl}
 & \bar{V}(k+1)-\bar{V}(k)\\
\le & -\frac{\beta}{2}  \displaystyle \sum_{i=1}^{N}\left(   \left(\frac{1}{2}-\beta d_{in}^{i}(k)  \left(\frac{l_{i}^{2}}{\alpha^{2}}+T_{s}\frac{l_{i}}{\alpha}\right) \right)\right.\sum_{j=1}^{N} a_{ij}(k) \cdot \\
&\left.  \displaystyle \|\Delta\hat{\lambda}_{j}(k) -\Delta\hat{\lambda}_{i}(k)\|^{2}  - \frac{d_{in}^{i}(k)\|e_i(k)\|^2}{\left(\frac{1}{2}-\beta d_{in}^{i}(k)  \left(\frac{l_{i}^{2}}{\alpha^{2}}+T_{s}\frac{l_{i}}{\alpha}\right) \right)}  \right)
\end{array}
\]
Suppose the condition  \eqref{eq:event-driven condition 2}
holds. Then it follows that  
\[
\begin{array}{rl}
   & \bar{V}(k+1)-\bar{V}(k) \\
\le &  -\frac{\beta}{2}(1-c_i) \sum_{i=1}^{N}  \left(\frac{1}{2}-\beta d_{in}^{i}(k)  \left(\frac{l_{i}^{2}}{\alpha^{2}}+T_{s}\frac{l_{i}}{\alpha}\right)\right)\\
 & \sum_{j=1}^{N} a_{ij}(k)\|\Delta\hat{\lambda}_{j}(k) -\Delta\hat{\lambda}_{i}(k)\|^{2}.
\end{array}
\]
Since $0<c_i<1$, it leads to $\bar{V}(k+1)-\bar{V}(k)\le 0$. 
Under Assumption \ref{assum 2}, the largest invariant set of $\{\Delta \hat{\lambda}|\bar{V}(k+1)-\bar{V}(k) =0\}$ is  $\{\Delta \hat{\lambda}|\Delta\hat{\lambda}_{1}=\ldots=\Delta\hat{\lambda}_{N}\}$. 
Therefore, according to the discrete-time LaSalle's invariance principle \cite{mei2017lasalle}, we have
that $\Delta\hat{\lambda}_{i}(k)-\Delta\hat{\lambda}_{j}(k)\rightarrow0,\forall i,j\in\mathcal{I}$
as $k\rightarrow\infty$. Then, it can be indicated from \eqref{eq:event-driven condition 2}
that $\lim_{k\rightarrow\infty}e_{i}(k)=\mathbf{0}$,
and hence, $\text{lim}_{k\rightarrow\infty}\Delta\bar{\lambda}_{i}(k)=\text{lim}_{t\rightarrow\infty}\Delta\hat{\lambda}_{i}(k),\forall i\in\mathcal{I}$.
It follows from \eqref{eq:algorithm-1} that $\lim_{t\rightarrow\infty}\dot{\gamma}=\mathbf{0}$.
}

\textcolor{black}{
Next, since the inequalities of \eqref{eq:beta_cond_sampling} and $c_i<1$
are strict, by following \eqref{eq:small perturbation} with similar
argument after \eqref{eq:small perturbation} in the proof of Theorem \ref{thm:convergence},
it can be proved that $\bar{V}(k+1)-\bar{V}(k)=0\Rightarrow\Delta\dot{\lambda}_{1}=\ldots=\Delta\dot{\lambda}_{N}=\mathbf{0}$.}

\textcolor{black}{Based on the result that $\text{lim}_{t\rightarrow\infty}\Delta\dot{\lambda}=\mathbf{0},\lim_{t\rightarrow\infty}\dot{\gamma}=\mathbf{0}$,
and $\text{lim}_{t\rightarrow\infty}\Delta\lambda=1_{N}\otimes s$
for some $s\in\mathbb{R}^{m}$, it can be concluded that the states
$\lambda$ and $\gamma$ under the algorithm \eqref{eq:algorithm-1}
with the triggering condition \eqref{eq:event-driven condition 2}
will converge to an equilibrium point $(\lambda^{*},\gamma^{*})$,
and $\lambda^{*}$ is identical to the optimal solution of \eqref{eq:distributed convex optimization}
if the initial condition satisfies $\sum_{i=1}^{N}\gamma_{i}(0)=\mathbf{0}$.}

\bibliographystyle{IEEEtran}
\bibliography{IEEEabrv,bibbysu}

\begin{thebibliography}{10}
\providecommand{\url}[1]{#1}
\csname url@samestyle\endcsname
\providecommand{\newblock}{\relax}
\providecommand{\bibinfo}[2]{#2}
\providecommand{\BIBentrySTDinterwordspacing}{\spaceskip=0pt\relax}
\providecommand{\BIBentryALTinterwordstretchfactor}{4}
\providecommand{\BIBentryALTinterwordspacing}{\spaceskip=\fontdimen2\font plus
\BIBentryALTinterwordstretchfactor\fontdimen3\font minus
  \fontdimen4\font\relax}
\providecommand{\BIBforeignlanguage}[2]{{%
\expandafter\ifx\csname l@#1\endcsname\relax
\typeout{** WARNING: IEEEtran.bst: No hyphenation pattern has been}%
\typeout{** loaded for the language `#1'. Using the pattern for}%
\typeout{** the default language instead.}%
\else
\language=\csname l@#1\endcsname
\fi
#2}}
\providecommand{\BIBdecl}{\relax}
\BIBdecl

\bibitem{nedic2009distributed}
A.~Nedic and A.~Ozdaglar, ``Distributed subgradient methods for multi-agent
  optimization,'' \emph{IEEE Transactions on Automatic Control}, vol.~54,
  no.~1, pp. 48--61, 2009.

\bibitem{zhu2011distributed}
M.~Zhu and S.~Mart{\'\i}nez, ``On distributed convex optimization under
  inequality and equality constraints,'' \emph{IEEE Transactions on Automatic
  Control}, vol.~57, no.~1, pp. 151--164, 2011.

\bibitem{gharesifard2013distributed}
B.~Gharesifard and J.~Cort{\'e}s, ``Distributed continuous-time convex
  optimization on weight-balanced digraphs,'' \emph{IEEE Transactions on
  Automatic Control}, vol.~59, no.~3, pp. 781--786, 2013.

\bibitem{shi2015extra}
W.~Shi, Q.~Ling, G.~Wu, and W.~Yin, ``Extra: An exact first-order algorithm for
  decentralized consensus optimization,'' \emph{SIAM Journal on Optimization},
  vol.~25, no.~2, pp. 944--966, 2015.

\bibitem{wang2011control}
J.~Wang and N.~Elia, ``A control perspective for centralized and distributed
  convex optimization,'' in \emph{2011 50th IEEE conference on decision and
  control and European control conference}.\hskip 1em plus 0.5em minus
  0.4em\relax IEEE, 2011, pp. 3800--3805.

\bibitem{yang2019survey}
T.~Yang, X.~Yi, J.~Wu, Y.~Yuan, D.~Wu, Z.~Meng, Y.~Hong, H.~Wang, Z.~Lin, and
  K.~H. Johansson, ``A survey of distributed optimization,'' \emph{Annual
  Reviews in Control}, vol.~47, pp. 278--305, 2019.

\bibitem{zhao2017distributed}
Y.~Zhao, Y.~Liu, G.~Wen, and G.~Chen, ``Distributed optimization for linear
  multiagent systems: Edge-and node-based adaptive designs,'' \emph{IEEE
  Transactions on Automatic Control}, vol.~62, no.~7, pp. 3602--3609, 2017.

\bibitem{cherukuri2015distributed}
A.~Cherukuri and J.~Cort{\'e}s, ``Distributed generator coordination for
  initialization and anytime optimization in economic dispatch,'' \emph{IEEE
  Transactions on Control of Network Systems}, vol.~2, no.~3, pp. 226--237,
  2015.

\bibitem{yi2016initialization}
P.~Yi, Y.~Hong, and F.~Liu, ``Initialization-free distributed algorithms for
  optimal resource allocation with feasibility constraints and application to
  economic dispatch of power systems,'' \emph{Automatica}, vol.~74, pp.
  259--269, 2016.

\bibitem{deng2017distributed}
Z.~Deng, S.~Liang, and Y.~Hong, ``Distributed continuous-time algorithms for
  resource allocation problems over weight-balanced digraphs,'' \emph{IEEE
  transactions on cybernetics}, vol.~48, no.~11, pp. 3116--3125, 2017.

\bibitem{kia2017distributed}
S.~S. Kia, ``Distributed optimal in-network resource allocation algorithm
  design via a control theoretic approach,'' \emph{Systems \& Control Letters},
  vol. 107, pp. 49--57, 2017.

\bibitem{ding2018distributed}
L.~Ding, G.~Y. Yin, W.~X. Zheng, Q.-L. Han \emph{et~al.}, ``Distributed energy
  management for smart grids with an event-triggered communication scheme,''
  \emph{IEEE Transactions on Control Systems Technology}, vol.~27, no.~5, pp.
  1950--1961, 2018.

\bibitem{zhu2019distributed}
Y.~Zhu, W.~Ren, W.~Yu, and G.~Wen, ``Distributed resource allocation over
  directed graphs via continuous-time algorithms,'' \emph{IEEE Transactions on
  Systems, Man, and Cybernetics: Systems}, 2019.

\bibitem{li2015distributed}
C.~Li, X.~Yu, W.~Yu, T.~Huang, and Z.-W. Liu, ``Distributed event-triggered
  scheme for economic dispatch in smart grids,'' \emph{IEEE Transactions on
  Industrial informatics}, vol.~12, no.~5, pp. 1775--1785, 2015.

\bibitem{shi2018distributed}
X.~Shi, Y.~Wang, S.~Song, and G.~Yan, ``Distributed optimisation for resource
  allocation with event-triggered communication over general directed
  topology,'' \emph{International Journal of Systems Science}, vol.~49, no.~6,
  pp. 1119--1130, 2018.

\bibitem{doan2020distributed}
T.~T. Doan and C.~L. Beck, ``Distributed resource allocation over dynamic
  networks with uncertainty,'' \emph{IEEE Transactions on Automatic Control},
  2020.

\bibitem{chopra2006passivity}
N.~Chopra and M.~W. Spong, ``Passivity-based control of multi-agent systems,''
  in \emph{Advances in robot control}.\hskip 1em plus 0.5em minus 0.4em\relax
  Springer, 2006, pp. 107--134.

\bibitem{li2019consensus}
M.~Li, L.~Su, and G.~Chesi, ``Consensus of heterogeneous multi-agent systems
  with diffusive couplings via passivity indices,'' \emph{IEEE Control Systems
  Letters}, vol.~3, no.~2, pp. 434--439, 2019.

\bibitem{yu2013output}
H.~Yu and P.~J. Antsaklis, ``Output synchronization of networked passive
  systems with event-driven communication,'' \emph{IEEE transactions on
  automatic control}, vol.~59, no.~3, pp. 750--756, 2013.

\bibitem{yan2019analysis}
Y.~Yan, L.~Su, V.~Gupta, and P.~Antsaklis, ``Analysis of two-dimensional
  feedback systems over networks using dissipativity,'' \emph{IEEE Transactions
  on Automatic Control}, 2019.

\bibitem{lee2014passivity}
P.~Lee, A.~Clark, L.~Bushnell, and R.~Poovendran, ``A passivity framework for
  modeling and mitigating wormhole attacks on networked control systems,''
  \emph{IEEE Transactions on Automatic Control}, vol.~59, no.~12, pp.
  3224--3237, 2014.

\bibitem{wang2014feedback}
Y.~Wang, M.~Xia, V.~Gupta, and P.~J. Antsaklis, ``On feedback passivity of
  discrete-time nonlinear networked control systems with packet drops,''
  \emph{IEEE Transactions on Automatic Control}, vol.~60, no.~9, pp.
  2434--2439, 2014.

\bibitem{zakeri2019recent}
H.~Zakeri and P.~J. Antsaklis, ``Recent advances in analysis and design of
  cyber-physical systems using passivity indices,'' in \emph{2019 27th
  Mediterranean Conference on Control and Automation (MED)}.\hskip 1em plus
  0.5em minus 0.4em\relax IEEE, 2019, pp. 31--36.

\bibitem{lu2017event}
Q.~L{\"u} and H.~Li, ``Event-triggered discrete-time distributed consensus
  optimization over time-varying graphs,'' \emph{Complexity}, vol. 2017, 2017.

\bibitem{kajiyama2018distributed}
Y.~Kajiyama, N.~Hayashi, and S.~Takai, ``Distributed subgradient method with
  edge-based event-triggered communication,'' \emph{IEEE Transactions on
  Automatic Control}, vol.~63, no.~7, pp. 2248--2255, 2018.

\bibitem{liu2019distributed}
C.~Liu, H.~Li, Y.~Shi, and D.~Xu, ``Distributed event-triggered gradient method
  for constrained convex minimization,'' \emph{IEEE Transactions on Automatic
  Control}, vol.~65, no.~2, pp. 778--785, 2019.

\bibitem{liu2020resource}
C.~Liu, H.~Li, and Y.~Shi, ``Resource-aware exact decentralized optimization
  using event-triggered broadcasting,'' \emph{IEEE Transactions on Automatic
  Control}, 2020.

\bibitem{kia2015distributed}
S.~S. Kia, J.~Cort{\'e}s, and S.~Mart{\'\i}nez, ``Distributed convex
  optimization via continuous-time coordination algorithms with discrete-time
  communication,'' \emph{Automatica}, vol.~55, pp. 254--264, 2015.

\bibitem{chen2016event}
W.~Chen and W.~Ren, ``Event-triggered zero-gradient-sum distributed consensus
  optimization over directed networks,'' \emph{Automatica}, vol.~65, pp.
  90--97, 2016.

\bibitem{yu2017distributed}
W.~Yu, Z.~Deng, H.~Zhou, and Y.~Hong, ``Distributed resource allocation
  optimization with discrete-time communication and application to economic
  dispatch in power systems,'' in \emph{13th IEEE Conference on Automation
  Science and Engineering (CASE)}.\hskip 1em plus 0.5em minus 0.4em\relax IEEE,
  2017, pp. 1226--1231.

\bibitem{du2018distributed}
W.~Du, X.~Yi, J.~George, K.~H. Johansson, and T.~Yang, ``Distributed
  optimization with dynamic event-triggered mechanisms,'' in \emph{2018 IEEE
  Conference on Decision and Control (CDC)}.\hskip 1em plus 0.5em minus
  0.4em\relax IEEE, 2018, pp. 969--974.

\bibitem{wang2018event}
A.~Wang, X.~Liao, and T.~Dong, ``Event-triggered gradient-based distributed
  optimisation for multi-agent systems with state consensus constraint,''
  \emph{IET Control Theory {\&} Applications}, vol.~12, no.~10, pp. 1515--1519,
  2018.

\bibitem{yi2018distributed}
X.~Yi, L.~Yao, T.~Yang, J.~George, and K.~H. Johansson, ``Distributed
  optimization for second-order multi-agent systems with dynamic
  event-triggered communication,'' in \emph{2018 IEEE Conference on Decision
  and Control (CDC)}.\hskip 1em plus 0.5em minus 0.4em\relax IEEE, 2018, pp.
  3397--3402.

\bibitem{liu2019event}
J.~Liu, W.~Chen, and H.~Dai, ``Event-triggered zero-gradient-sum distributed
  convex optimisation over networks with time-varying topologies,''
  \emph{International Journal of Control}, vol.~92, no.~12, pp. 2829--2841,
  2019.

\bibitem{shi2020distributed}
X.~Shi, Z.~Lin, T.~Yang, and X.~Wang, ``Distributed dynamic event-triggered
  algorithm with minimum inter-event time for multi-agent convex
  optimisation,'' \emph{International Journal of Systems Science}, pp. 1--12,
  2020.

\bibitem{liu2016event}
S.~Liu, L.~Xie, and D.~E. Quevedo, ``Event-triggered quantized
  communication-based distributed convex optimization,'' \emph{IEEE
  Transactions on Control of Network Systems}, vol.~5, no.~1, pp. 167--178,
  2016.

\bibitem{tang2016distributed}
Y.~Tang, Y.~Hong, and P.~Yi, ``Distributed optimization design based on
  passivity technique,'' in \emph{2016 12th IEEE International Conference on
  Control and Automation (ICCA)}.\hskip 1em plus 0.5em minus 0.4em\relax IEEE,
  2016, pp. 732--737.

\bibitem{hatanaka2018passivity}
T.~Hatanaka, N.~Chopra, T.~Ishizaki, and N.~Li, ``Passivity-based distributed
  optimization with communication delays using pi consensus algorithm,''
  \emph{IEEE Transactions on Automatic Control}, vol.~63, no.~12, pp.
  4421--4428, 2018.

\bibitem{bao2007process}
J.~Bao and P.~L. Lee, \emph{Process control: the passive systems
  approach}.\hskip 1em plus 0.5em minus 0.4em\relax Springer Science \&
  Business Media, 2007.

\bibitem{minty1964monotonicity}
G.~J. Minty \emph{et~al.}, ``On the monotonicity of the gradient of a convex
  function.'' \emph{Pacific Journal of Mathematics}, vol.~14, no.~1, pp.
  243--247, 1964.

\bibitem{bertsekas1996neuro}
D.~P. Bertsekas and J.~N. Tsitsiklis, \emph{Neuro-dynamic programming}.\hskip
  1em plus 0.5em minus 0.4em\relax Athena Scientific Belmont, MA, 1996, vol.~5.

\bibitem{ruszczynski2006nonlinear}
A.~Ruszczy{\'n}ski, \emph{Nonlinear optimization}.\hskip 1em plus 0.5em minus
  0.4em\relax Princeton university press, 2006, vol.~13.

\bibitem{wan2009event}
P.~Wan and M.~D. Lemmon, ``Event-triggered distributed optimization in sensor
  networks,'' in \emph{Proceedings of the 2009 International Conference on
  Information Processing in Sensor Networks}.\hskip 1em plus 0.5em minus
  0.4em\relax IEEE Computer Society, 2009, pp. 49--60.

\bibitem{mei2017lasalle}
W.~Mei and F.~Bullo, ``La{S}alle invariance principle for discrete-time
  dynamical systems: A concise and self-contained tutorial,'' \emph{arXiv
  preprint arXiv:1710.03710}, 2017.

\end{thebibliography}

\end{document}